\newcommand{\cappa}{ {\cal K} }   		
\newcommand{\cappaG}{ {\cal K}_{\text{\tiny{\it Gauss}}}   }	
\newcommand{\om}{\omega}			
\newcommand{\Om}{\Omega}			
\newcommand{\Oms}{\Omega_s}	
\newcommand{\Omi}{\Omega_i}

\newcommand{\DDpm}{{\cal D}^{\hspace{-0pt}\text{\tiny{$\pm$}}}}

\newcommand{\bt}{\bar{t}}
\newcommand{\bts}{\bar{t}_s}
\newcommand{\bti}{\bar{t}_i}
\newcommand{\tcorr}{ (\delta t_s )_{cond.} }
\newcommand{\Dtau}{\Delta \tau}
\newcommand{\spj}{}
\newcommand{\spa}{}

\newcommand{\sppm}{}
\newcommand{\taus}{ {\tau_{s}} }
\newcommand{\taui}{{\tau_{i}^{\text{\tiny{$+$}}}}}

\newcommand{\tauib}{{\tau_{i}^{\text{\tiny{$-$}}}}}

\newcommand{\tauipm}{{\tau_{i}^{\text{\tiny{$\pm$}}}}}
\newcommand{\tauispm}{ \tauipm-\taus}

\newcommand{\taup}{ { T}_p}
\newcommand{\gamgauss}{\gamma}

\newcommand{\as}{\hat{a}_s}
\newcommand{\ai}{\hat{a}_i}

\newcommand{\sinc}{{\rm sinc}}

\newcommand{\nn}{\nonumber}
\newcommand{\bsub}{\begin{subequations}}
\newcommand{\esub}{\end{subequations}}
\newcommand{\beq}{\begin{equation}}
\newcommand{\eeq}{\end{equation}}
\newcommand{\beqa}{\begin{eqnarray}}
\newcommand{\eeqa}{\end{eqnarray}}
\newcommand{\beql}{\begin{subequations}\begin{eqnarray}}
\newcommand{\eeql}{\end{eqnarray}\end{subequations}}
\documentclass[pra,aps,amsmath, nofootinbib,showpacs]{revtex4-1}
\usepackage{amsmath} 
\usepackage{empheq}
\usepackage{amssymb}
\usepackage{graphicx}
\usepackage{longtable}
\usepackage{float}   
\usepackage{verbatim}  
\usepackage{braket}
\usepackage{color}
\begin{document}
\title{Heralding pure single photons: a comparision between counter-propagating and co-propagating twin photons}
\author{ Alessandra~Gatti$^{1,2}$ and Enrico~Brambilla$^2$  }
\affiliation{$^1$ Istituto di Fotonica e Nanotecnologie del CNR, Piazza Leonardo  Da Vinci 32, Milano, Italy;  
$^2$ Dipartimento di Scienza e Alta Tecnologia dell' Universit\`a dell'Insubria, Via Valleggio 11,  Como, Italy}
\email{Alessandra.Gatti@mi.infn.it}
\begin{abstract}
We investigate different strategies suitable to generate  pure heralded single photons through spontaneous parametric down-conversion, comparing  the counter-propagating geometry studied in \cite{gatti2015} with more conventional co-propagating configurations which enhance the purity of the heralded photon state through the technique of group-velocity matching. 
Our analysis is based on the correlation of twin photons in the temporal domain,a non-standard approach that  provides a physical view of the mechanisms that permit to eliminate  the temporal entanglement of the state and to generate high-purity heralded photons. 
The Schmidt number associated to the temporal modes, which provides  a more quantitative estimate of the purity, is then  calculated. The  efficiency of the various strategies and the individual properties of the heralded photons thereby generated are also compared. 
\end{abstract}
\maketitle
\section*{Introduction}
\label{sec:intro}
\begin{figure}[h]
\includegraphics[scale=0.38]{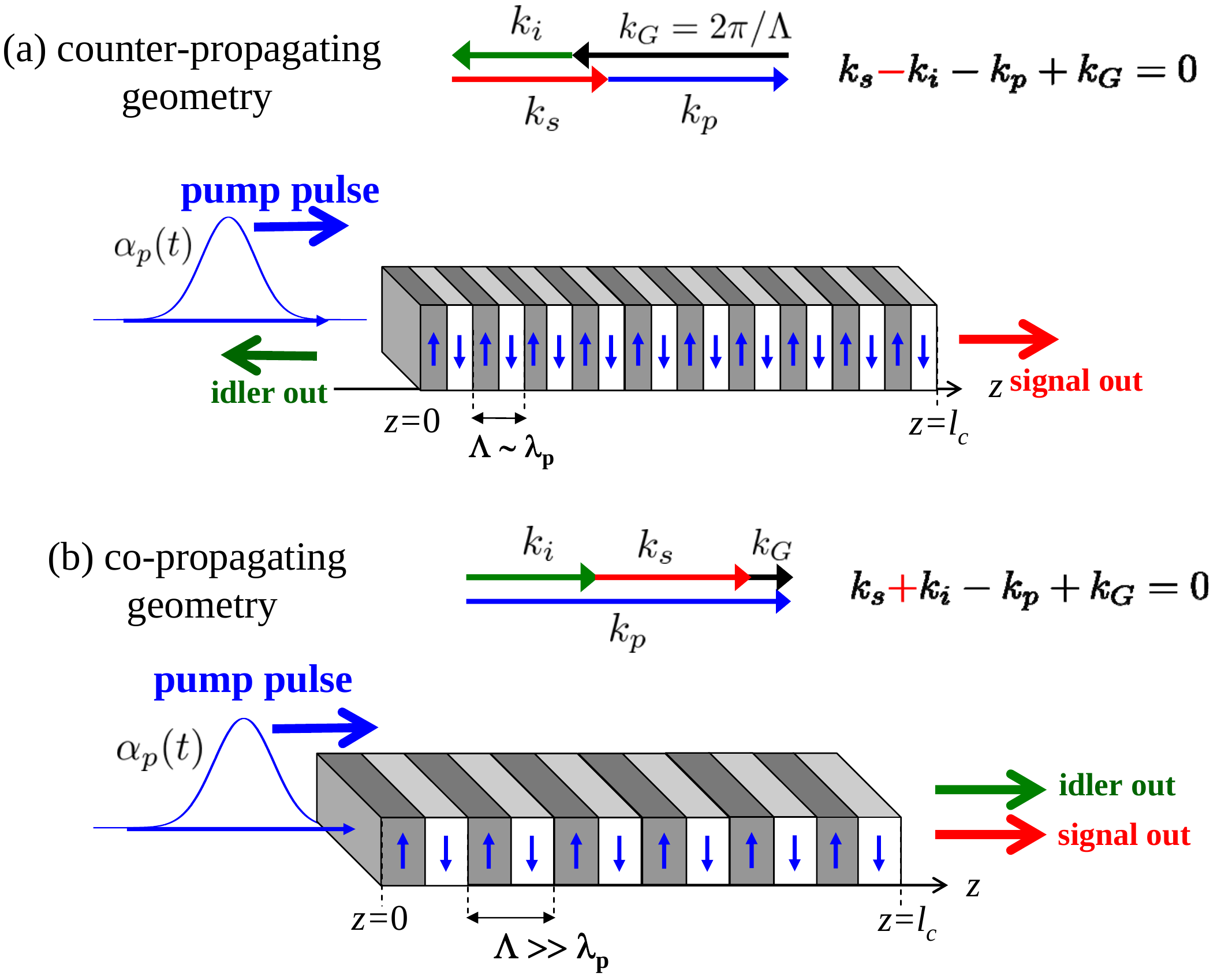}
\caption{(a) Scheme of  twin-photon  generation  in the (a)  counter-propagating and (b) 
co-propagating geometries. 
In the case (a)  quasi-phasematching at first order requires submicrometer
poling periods $\Lambda \sim \lambda_p $  of the nonlinear material
.}
\label{fig1}
\end{figure}

Single-photon states  
are of outstanding interest in modern quantum optics, as the basis of fundamental tests of quantum mechanics and  of a variety  of quantum technologies
(see e.g. \cite{migdall2011, migdall2013} for recent surveys of single-photon  techniques and applications).  In order to generate truly single photons, 
one of the most efficient method is based  
on a conditional measurement, where a  two-photon state  is generated and the presence of a single photon is  {\em heralded}  by detection of   its partner.  
To this end,   parametric  processes such as spontaneous   parametric down-conversion (PDC) or four-wave mixing (FWM), where one or two photons belonging to a high energy pump laser are occasionally converted  into pairs of photons, are routinely employed. The conservation laws ruling these microscopic  processes originate a quantum correlation in the spatial and temporal degrees of freedom of the pair, which typically extends  over broad spectral and angular bandwidths. Such a {high dimensional  entanglement} may represent  a resource for  broadband quantum communication or quantum metrology, but is detrimental for the purity of  heralded single photons, because detection of the trigger photon projects the state of its twin in a highly mixed state. This represents  a  limitation for quantum communication/information protocols where single photons are required to be in indistinguishable and capable of high-visibility interference. \\
The recent development of waveguided PDC (see \cite{christ2013}  and reference therein) and of FWM in single-mode fibers  \cite{mcmillan2013} opened the possibility to control  the spatial degrees of freedom, and to generate the twin photons into a single of few spatial modes. In order to eliminate also  the spectro-temporal correlation,  a possibility  is filtering  hard enough  that a single spectral mode is selected, but clearly the efficiency of the source is reduced. 
In order to achieve pure heralded photons with high fluxes, therefore,  considerable effort has been devoted  
to find alternative techniques,  which directly ruduce the degree of entanglement of the source in order to produce
uncorrelated twin photons\cite{grice2001,uren2006,mosley2008b, mosley2008, migdall2002,levine2010,bennink2010,branczyk2011,zhang2012}.
In such
a way a conditional
measurement projects the field in a pure single-photon state rather than in a mixed state.    In the standard co-propagating  geometry (Fig.\ref{fig1}b) this task requires careful techniques of group velocity matching, which can be implemented only in some materials and tuning conditions\cite{grice2001, mosley2008}. 
\par
In recent years,  the counter-propagating configuration of PDC, where twin photons are emitted along opposite directions (Fig.\ref{fig1}a),  emerged  as a promising source of  heralded pure single photons  without the need  of  group-velocity matching \cite{christ2009,gatti2015}. 
 Proposed by Harris in the sixties \cite{harris66} and implemented in 2007 by Canalias et al. \cite{canalias2007}, this configuration presents the challenge of requiring a very short  poling of the nonlinear material\cite{canalias2003,canalias2012}.  On the other hand, counter-propagating twin photons and twin beams  possess peculiar and attractive features, such as their narrowband character \cite{canalias2012b,suhara2010,gatti2015,corti2016} and the potentiality to generate in the high-gain regime a robust continuous-variable entanglement  \cite{gatti2017}. In the spontaneus regime, twin-photons may be naturally generated in either spectrally entangled or decorrelated states by simply modifying the pump pulse duration (or crystal length), without the need of special tuning conditions \cite{gatti2015}.  Similar features have recently been described in counterpropagating FWM \cite{uren2016}.
\par 
In this work we focus on the  PDC process, and on the temporal correlation of twin photons thereby generated, making  a parallel analysis of the two co-propagating and counterpropagating geometries  (Fig.\ref{fig1}) as sources of
heralded single photons. 
By a systematic comparison of the two configurations,  we aim at providing a deeper understanding of the physical mechanisms under which temporal correlation emerges and can eventually be eliminated, 
which may turn useful to optimize the existing configurations and to design new ones. 
Event though our analysis is restricted to the PDC process, some of our results may  qualitatively hold also for FWM. 
\par
A large part of our analysis will be based on the  correlation of twin-photons in the temporal domain. This approach  is not standard in the literature\cite{christ2009, grice2001,uren2006,mosley2008b, mosley2008, migdall2002,levine2010,bennink2010,branczyk2011,zhang2012}, which mostly focused on their spectral correlation, and provides, in our opinion, a more direct physical view. For example, it will show that the  
temporal correlation between twin photons can be eliminated only creating conditions such that the timing provided by the pump pulse is more precise than that offered by detecting any of the twin photons. In this way,  detection of  one photon does not give any better information on the exit time of its twin than by not detecting it, and the two-photon state appears uncorrelated. \\
From a more quantitative point of view,  we  will derive  an approximated  formula for the Schmidt number, to our knowledge not present in  the former literature, which holds for both geometries and gives a simple means  for evaluating the degree of purity of the heralded photon based only on two parameters, that describe the characteristic time scales of the source.\\
Finally,  the individual spectro-temporal properties of the heralded photons, which are  of great importance in view of different applications will be investigated in parallel.  In particular, we will see  that either narrowband or broadband pure heralded photons are characteristic of the two geometries, and that different strategies of group-velocity matching lead to very different efficiencies of pair production. 
\par 
The paper is organized as follows: 
After introducing in Sec.\ref{sec:phasematching} the two geometries and their  different phase-matching conditions,  the core of our results is contained in sections \ref{sec:temporal}-\ref{sec:spectrum}.   Sec.\ref{sec:temporal} describes the temporal correlation of twin photons, in terms both of general formulas valid for any configuration of PDC, and of numerical calculations for  specific examples. 
Sections \ref{sec:Schmidt} calculates the Schmidt number of entanglement. General analytic results obtained within the Gaussian approximation for the biphoton amplitude are compared with more exact numerical ones. 
Sec \ref{sec:interpretation} offers a physical interpretation of the results obtained, while the final  Sec.\ref{sec:spectrum} describes the spectro-temporal properties of heralded photons and compares the efficiency of heralding within the different strategies investigated. 

\section{Counter-propagating  and co-propagating geometries}
\label{sec:phasematching}
As in most of the previous literature, our analysis is restricted to the purely temporal domain,  assuming either a single-mode   waveguided configuration, or that 
a small angular bandwidth is collected. 
\par
In the counter-propagating geometry shown in Fig.\ref{fig1}a, 
a coherent pump pulse of central frequency $\omega_p$ and temporal profile $\alpha_p(t)$ impinges a nonlinear $\chi^{(2)}$ crystal 
of length $l_c$ from the left face, 
and generates counter-propagating photon pairs with, say,  the idler photon  backpropagating towards the laser source.  For the counter-propagating three-wave interaction, momentum conservation  requires quasi-phase matching in a  periodically poled structure, with a short poling  period $\Lambda$,  such that  the momentum associated to the nonlinear grating of  the nonlinearity $k_G=\frac{2  \pi m}{\Lambda}$, $m=\pm 1, \pm3\hdots$,   approximately compensates  the momentum of the pump photon.  At first order,  it requires  $\Lambda$  on the same order as the pump wavelength in the medium. 
 The central frequencies of the emitted signal
and idler fields, $\omega_s$ and $\omega_i=\omega_p-\omega_s$, 
are thus determined by the  poling period $\Lambda$ and the pump central frequency $\omega_p$
according to
\beq
\label{qpm}
k_{s}-k_{i}=k_{p}-k_G\qquad \text{\small{{\bf  (a)counter-propagating case }}}
\eeq
where $k_{j}=\frac{\om_j}{c}n_j(\omega_j)$, $j=s,i,p$ are the wave-numbers
at the corresponding central frequencies $\omega_j$.
\par
In the co-propagating geometry (Fig.\ref{fig1}b)  all the three   fields propagate  along the positive $z$ direction. In this case momentum conservation requires 
\beq
\label{qpm2}
k_{s}+k_{i}=k_{p}-k_G\qquad\text{\small{{\bf (b) co-propagating case}}}
\eeq
which  describes   both  the case of a bulk crystal, in which   $k_G=0$,  or quasi-phasematching in periodically poled
structures.

 The efficiency of  the down-conversion process is described by a dimensionless gain parameter  $g$, proportional to the peak  amplitude of the pump laser, the crystal length and the nonlinear susceptibility. Considering the spontaneous regime  where $g \ll 1$, and retaining terms at most linear in $g\ll1$, 
the signal-idler state  at the crystal output  takes  the well-known form 
\begin{align}
&|\phi^{\spj}\rangle = \left |0\right \rangle  + |\phi^{\spj}_2 \rangle\label{state}\\
&|\phi^{\spj}_2\rangle= \int d\Om_s d\Om_i \psi^{\sppm} (\Om_s, \Om_i) \as^\dagger (\Om_s) \ai^\dagger (\Om_i) \left |0\right \rangle \, ,
\label{statec}
\end{align}
corresponding to the superposition of the vacuum  and the  two-photon state $|\phi^{\spj}_2\rangle$ (see e.g.  \cite{gatti2015} for a derivation from the more powerful quantum-field description). Here
$\as^\dagger (\Om_s)$ and $\ai^\dagger (\Om_i)$  are   signal and idler photon creation operators in the frequency domain,  $\Om_j$ being the offset from the reference frequency $\omega_j$, and 
\beq
\label{psi}
\psi^{\sppm}(\Omega_s,\Omega_i)
=\frac{g }{\sqrt{2\pi}}   
\tilde{\alpha}_p(\Omega_s+\Omega_i)   
\mathrm{sinc}\left[\frac{\DDpm(\Omega_s,\Omega_i)l_c}{2}\right]    e^{i\beta(\Omega_s,\Omega_i) }
\eeq
is the so-called spectral biphoton amplitude, giving 
the joint probability amplitude of detecting a signal photon at frequency $\omega_s+\Om_s$
and an idler photon at frequency $\omega_i+\Om_i$. In formula \eqref{psi}
\beq
\label{alphap}
\tilde{\alpha}_p(\Om)=\int\frac{dt}{\sqrt{2\pi}}e^{i\Om t} {\alpha_p(t)}\, ; 
\eeq
 is the  spectral amplitude of the pump pulse, 
where the  temporal profile ${\alpha_p(t)}$  is  normalized to its peak value;  
$\DDpm(\Oms,\Omi)l_c$  is the phase-mismatch function, where here and in the following 
the index  $+ $  and $-$ refer to the counter-propagating  and co-propagating geometries (a) and (b), respectively 
\beqa
\DDpm(\Oms,\Omi) 
&&=
\left\{
\begin{array}{lr}
 k_p(\Oms+\Omi)  -k_s(\Oms) + k_i(\Omi) - k_G&\quad \text{  \; \bf (a)}
\vspace{3pt}\\
k_p(\Oms+\Omi)- k_s(\Oms)-k_i(\Omi)- k_G&\quad \text{ \; \bf (b)}
\end{array}
\right.
\label{mismatch}
\eeqa
Finally
\beq
\beta (\Oms,\Omi)= \frac{l_c}{2} \left[ k_s(\Oms) +k_i (\Omi) + k_p (\Omega_s + \Omega_i) \right]
\eeq
is a global phase,   acquired during the propagation along the crystal in the absence of any nonlinear effect  by a pair of photons generated at the crystal center $z=l_c/2$. \\
As described in \cite{gatti2015, corti2016,gatti2017},  the different sign in front of the idler wave-number $k_i(\Om_i)$  in Eq. \eqref {mismatch}   is at the origin  of the radically different  properties   of the two geometries of PDC.  
This is best seen by expanding the phase mismatch (\ref{mismatch}) at first order around the reference frequencies (corresponding to $\Omega_j=0$), 
\beqa
\label{Dlin}
\DDpm (\Oms,\Omi)  \frac{l_c}{2}  
&\approx&\frac{l_c}{2}[(k_p'\pm  k_i')\Omi +(k_p'- k_s')\Oms   ]       \nn  \\
&\equiv&   \tauipm\Omi + \taus\Oms \;,\label{Dlin2}
\eeqa
where 
$k_j'\equiv v_{gj}^{-1} =\left(\frac{d k_j}{d \Omega_j}\right)_{\Omega_j=0}$
is the inverse group velocity of wave $j$  and we introduced the  characteristic times
\beqa
\taus = \frac{1}{2}\left(\frac{l_c}{v_{gp}} - \frac{l_c}{v_{gs}}\right) \label{taus}\\
\tauipm= \frac{1}{2}\left(\frac{l_c}{v_{gp}}\pm\frac{l_c}{v_{gi}}\right) \label{taui}
\eeqa
which 
involve either the difference or the sum of the  group velocities  
of the pump  and the  down-converted  field, 
depending whether the latter co-propagates or counterpropagates with respect to the former, 
 or, in other words,  they  involve the {\em relative} group  velocities of the down-converted photons with respect to the pump. These constants represent   the characteristic   temporal separations between the centers of the pump and down-converted wave-packets,  and describe how much delayed from the pump  a pair of twin photons generated at the crystal center arrive at their end faces. 
In the co-propagating  case  $\taus$ and $\tauib$ are both determined by the group velocity mismatch (GVM)  with respect to the  pump,  and are typically on the same order of magnitude,  unless some particular strategy of group velocity matching is employed. 
By contrast, in the counter-propagating case, the time constant associated to the  backward photon
$\taui$  involves  the
group velocities sum (GVS) and  is  on the order of the photon transit time across the crystal, which
exceeds  the GVM  time  by one or two orders of magnitude. 
\\
As will be clear in the following (see also \cite{grice2001,uren2006,grice2010,gatti2015}), 
the possibility to generate
heralded photons with a high degree of purity depends on the relative sizes and signs of these  time constants,  and on how they compare
to the pump duration $\taup$.   Therefore we introduce    the  ratio  
\begin{equation}
\eta^{\spa}=\frac{\taus}{\tauipm} \;,
\label{eta}
\end{equation}
which can be positive or negative ( only $\taui$  is always positive, while $\taus$ and $\tauib$ can be either positive or negative). 
Clearly,  
in  the counter-propagating case  \beq
 |\eta| = \frac {|k'_p -k'_s|}{k'_p +k'_i} =\frac{|\taus| }{\taui}  \ll 1 
\eeq 
for any choice of materials and phase matching conditions. 
Without loss of generality, we assume  that in the co-propagating case the signal and idler are chosen  so  that  $|\taus|\leq|\tauib|$, 
so that in both configurations
\beq
-1\leq\eta^{\sppm}\leq 1
\label{etarange}
\eeq

An analogous linear approximation  of the global phase appearing in Eq.\eqref{psi} gives: 
\begin{align}
\beta (\Oms,\Omi) &\approx \frac{l_c}{2} \left[ k_s  +k_i  + k_p   + (k_s' + k'_p) \Oms
+ (k_i'  + k'_p ) \Omega_i\right] \nn \\
&= \text{const.} +  t_{As}   \Oms +  t_{Ai}   \Omi \, ,
\label{betalin} 
\end{align} 
where 
\bsub
\label{tA}
\begin{align} 
t_{As} &= \frac{l_c}{2v_{gs}}+ \frac{l_c}{2v_{gp}}    \label{tAs} = t_{Ap} - \taus\\
t_{Ai} &=   \frac{l_c}{2v_{gi}}+ \frac{l_c}{2v_{gp}}  \label{tAi}= t_{Ap} -\tauib
\end {align}
\esub
can be considered as  the times at which two twin photons downconverted from the pump peak at  $z= l_c/2$ -  pictured as  wave-packets propagating without deformation - arrive at their respective end faces of the crystal. They 
represent the exit time of the centers of the signal and idler wavepackets, assuming that at   $t=0$ the center of the pump pulse enters  the crystal ($t_{Ap} = \frac{l_c}{v_{gp}}   $  is then the exit time of the pump) . 

We notice that the omission of second and higher order terms in Eqs. \eqref{Dlin2} and \eqref{betalin} is valid only for small bandwidths, and corresponds to neglecting the temporal dispersion. This is 
is  well justified in the counter-propagating configuration,  which involves narrow down-conversion spectra \cite{canalias2007,gatti2015, corti2016}, while it is less justified in the  co-propagating case,   because of the larger bandwidths in play. In particular, it is not justified when  $\taus \simeq\tauib$ ( e.g. for type I PDC close to degeneracy). 

\section{Time-domain view: the temporal correlation of twin photons}
\label{sec:temporal}
An alternative 
and  perhaps more intuitive 
view of the problem is offered by  the joint amplitude in the temporal domain, which  is obtained
by back-Fourier transforming  the joint spectral amplitude
\beq
\label{phi}
\phi(t_s,t_i)=\int\frac{d\Oms}{\sqrt{2\pi}}
\int\frac{d\Omi}{\sqrt{2\pi}}
e^{-i (\Oms t_s+\Omi t_i)} \psi(\Oms,\Omi)
\eeq
In terms of this function, the two-photon state \eqref{statec} becomes
\begin{align}
&|\phi^{\spj}_2 \rangle= \int d t_s d t_i \phi (t_s,  t_i) \as^\dagger (t_s) \ai^\dagger (t_i) \left |0\right \rangle \, ,
\label{statec2}
\end{align}
where $ \as^\dagger (t_s)$, $\ai^\dagger (t_i) $ are photon creation operators in the time domain, and for example 
$\as^\dagger (t_s) |0\rangle \equiv |t_s\rangle $  represents the state with exactly one signal photon at time $t_s$ at the crystal output face. Thus,  $\phi(t_s,t_i)$, 
which was extensively  analysed in \cite{gatti2015} in  the counter-propagating case, represents the joint probability amplitude that a signal and an idler photon exit the crystal slab at times  $t_s$ and $ t_i$, respectively, and describes the temporal correlation between the twin photons. The rate of coincidence counts  at the crystal exit faces is then given by 
\begin{align} 
G_{si}^{(2)} (t_s,t_i) \equiv \langle  \as^{\dagger} (t_s) \as  (t_s) \ai^{\dagger} (t_i) \ai  (t_i) 
\rangle \approx \left| \phi (t_s,t_i)\right|^2 .
\label{G2}
\end{align}
\par
\begin{figure*} 
\centering
\includegraphics[scale=0.65]{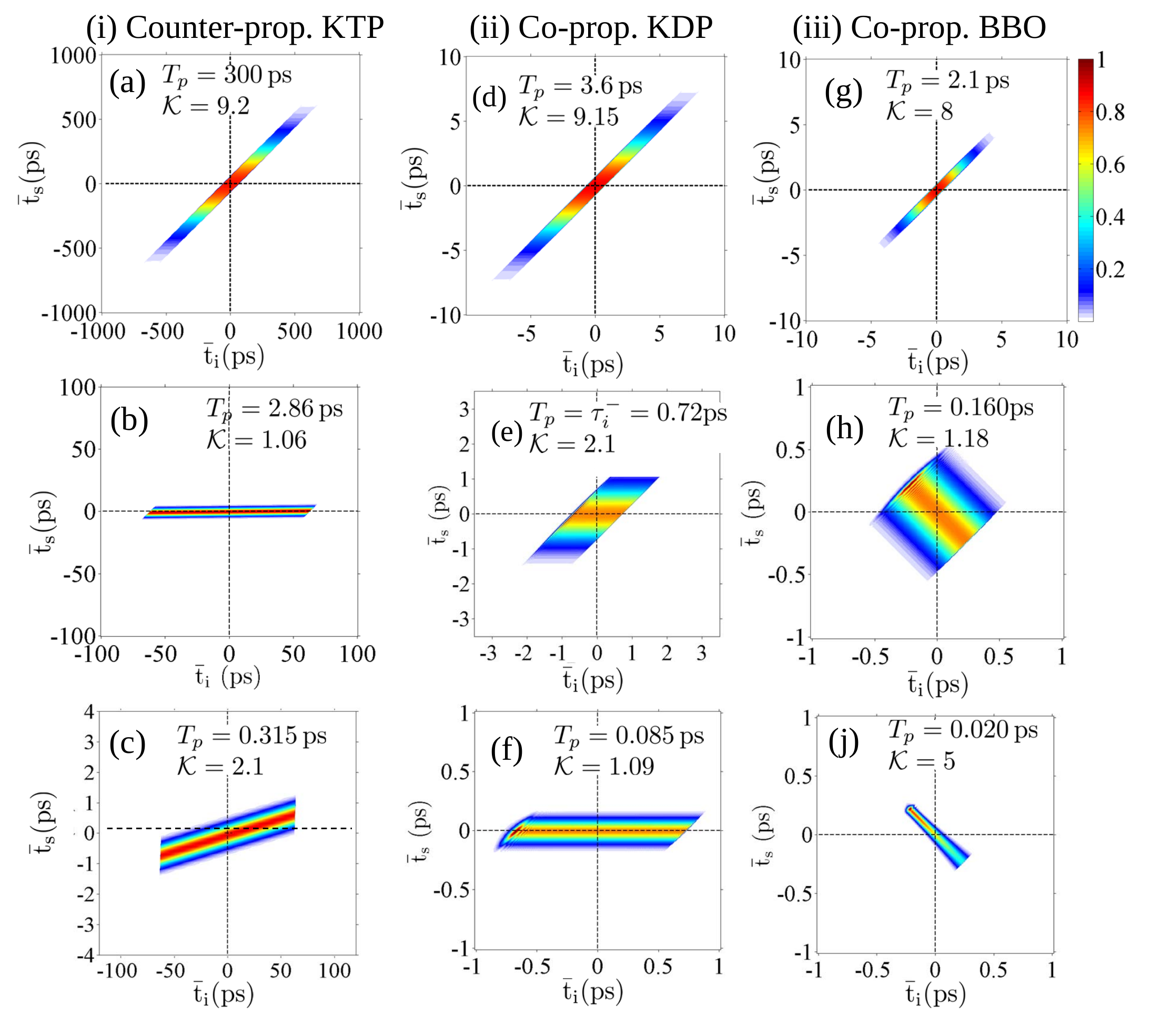}
\caption{Temporal correlation $|\phi(\bar t_s,\bar t_i)|^2$, representing the joint  probability of finding a signal and an idler twin photons at times $t_s = \bts + t_{As}$ and $t_i = \bts + t_{Ai}$ at their crystal exit faces, where $ t_{Aj}$ are the reference times defined by Eq.\eqref{tA}. In each panel $\cappa$ shows the Schmidt number of entanglement (see next section), with $\cappa=1$ corresponding to perfect separability. 
 In each column the pump  duration $\taup$ decreases from top to bottom, and
panels (b), (f), (h) correspond  to  the minima  of $\cappa$ in Fig.\ref{fig3}, and represent  the most  separable conditions for each example} 
\label{fig4_temp}
\end{figure*}

 
Several plots of this correlation function are shown in Fig.\ref{fig4_temp},  where the three columns correspond to three examples chosen as representatives of the two geometries  (see Table \ref{table1} for parameters, and Appendix \ref{sec:examples} for details), namely {\bf (i)} a generic counter-propagating configuration, not specifically optimized for separability, where $\eta\simeq 0.01$, and two co-propagating configurations optimized for separability, with {\bf (ii)}  $\eta=0$, and {\bf (iii)} $\eta=-1$. 
Condition {(ii)}  is referred to as {\em asymmetric group-velocity matching}  and  requires that the signal photon propagates with the same group velocity as the pump ($\taus=0$)\cite{uren2006, mosley2008b}. Condition { (iii)}  corresponds to the {\em symmetric group-velocity matching}, and can be realized only in the co-propagating case, requiring $\taus=-\taui$\cite{grice2001, uren2006}. 
They are usually difficult to satisfy in the visible range,  but can be achieved in some  $\chi^{(2)}$ material in the near infrared and at telecom wavelengths \cite{grice2001,uren2006, mosley2008b, mosley2008, migdall2013}. Experimental evidence of frequency decorrelated photon pairs through this technique
were first reported in \cite{mosley2008b}.
\begin{table*}
\begin{tabular}{|l|c|c|c|c|c|c|c|c|c|}
 \hline\hline
\textbf{crystal}  &$l_c$ (mm) & phase matching ($\theta_p$)  & $\lambda_p$ &$\lambda_s\,$&$\lambda_i\,$&$\taus$\,  
                    & $\tauipm$ & $\taup^{min}$ & $\eta$\\ \hline
{\bf (i)} PPKTP  & 10mm& type 0  e-ee ($90^\circ$) &821.4nm&1141nm& 2932nm  & 0.67ps & 63ps &4.05ps &0.01\\
{\bf (ii)} KDP  & 10mm& type II e-oe ($67.8^\circ$) & 415nm &830nm& 830nm  &  0     & 0.72ps & 0     & 0\\
{\bf (iii)} BBO & 10mm& type II e-oe ($28.8^\circ$) & 757nm &1514nm& 1514nm &  -0.237ps & 0.237ps & 0.147ps &-1 \\
\hline\hline
\end{tabular}
\caption{Phase-matching conditions and characteristic time constants  for the three crystals taken as examples: 
(i) periodically poled KTP, with $800$nm poling period for the counter-propagating configuration, 
(ii)  KDP and (iii)  BBO bulk crystal for the two co-propagating configurations.}\label{table1}
\end{table*}
\par 
The plots in Fig.\ref {fig4_temp} have been calculated from Eq.\eqref{phi} without resorting to any approximation,  by using the complete Sellmeier dispersion formula in \cite{niko91,kato2002,zernike64} for evaluating the wave-numbers.  If the linear approximations \eqref{Dlin2} and \eqref{betalin} are instead employed, the Fourier transform (\ref{phi})  can  be explicitly calculated  \cite{gatti2015}, obtaining 
\begin{align} 
\phi(\bt_s,\bt_i) =
\frac{g}{2|\tauispm|}
{\alpha}_p\left(\frac{\bar{t}_s-\eta\bar{t}_i}{1-\eta}\right)
{\rm Rect}\left(\frac{\bar{t}_s-\bar{t}_i}{|\tauipm -\taus|}\right)
\label{psitemp}
\end{align}
where constant phase factors have been omitted. 
The  barred arguments 
\beqa
\bar{t}_j&=&t_j- t_{Aj}   \qquad j=s,i   \, ,
\eeqa
are the exit times of the twin photons measured  relative to the central  exit times $t_{Aj}$ of their wavepackets, 
 [see  Eq. \eqref{tA}]. The temporal correlation in Eq.\eqref{psitemp} is the product of two factors:  the first one is 
the pump temporal profile ${\alpha}_p(t)$, that we can e.g. take as a Gaussian, 
the second one 
is  the box function
\beq
{\rm Rect}\left(\frac{t}{\Dtau } \right)=\begin{cases}  1  \qquad &\text{for } \, t\epsilon \, [-\Dtau, \Dtau]\\
0 \qquad &\text{elsewhere}
\end{cases}  
\eeq
 of  width $2\Dtau$ and unitary height, where 
\beq
\Dtau:=|\tauipm-\taus|=\frac{l_c}{2}\left|\frac{1}{v_{gs}}\pm\frac{1}{v_{gi}}\right|
\label{Dtau}
\eeq
represents the typical delay  intercurring between the exit times of two twin photons. 
\par 
The result \eqref{psitemp} relies
 merely on linearization of the propagation phases, and as such fails to be valid for $\taus \to \tauipm$, when the first order of the Taylor espansion of $\DDpm$  vanishes. 
 It  coincides with the expression analysed  in  \cite{gatti2015}, where it was shown that  in the counter-propagating configuration   the temporal correlation \eqref{psitemp}  reduces to  a  factorable function of $\bar t_s, \bar t_i$   for a range of pump durations  intermediate between the two  characteristic time scales:  $|\taui| \gg  \taup\gg |\taus|$. The analysis  in \cite{gatti2015} can then be generalized to a generic geometry.  
\par 
From a geometric/mathematic viewpoint, we have the noticeable asymptotic behaviors of the temporal correlation in Eq.\eqref{psitemp} : \\
 {\bf 1)}Limit of a long pump pulse $  {\taup  \gg |\tauipm|\ge |\taus|}$. 
When the pump pulse is much longer than both characteristic  time scales, the pump profile has a slow variation with respect to the  much narrower box function. Therefore we can put  $\bar t_s= \bar t_i$ inside the term 
${\alpha}_p\left(\frac{\bar{t}_s-\eta\bar{t}_i}{1-\eta}\right)$, and
\begin{align}
\phi(\bt_s,\bt_i) \underset{ \taup \gg |\tauipm|} {\longrightarrow}  &
\frac{g}{2|\tauispm|}
\alpha_p \left(\bar{t}_s \right)
{\rm Rect}\left(      \frac{\bar{t}_s-\bar t_i }{|\tauipm -\taus|}       \right])
\label{asym1}
\end{align}
This formula  well reproduces  the three upper plots  in  Fig.\ref{fig4_temp}, which exhibit  a a sharp maximum of the temporal correlation along the diagonal  $\bar t_s = \bar t_i$, espressing the fact that twin photons can be generated at any time along the pump pulse, but they tends to exit almost simultaneously from the crystal (a part from the fixed offset $t_{As}-t_{Ai}$)   with a  flat distribution of their mutual time delay in the interval $[-\Dtau,  \Dtau]$. 
This situation corresponds to a  joint temporal amplitude not factorable in its arguments, and the two-photon  state  \eqref{statec2} is  entangled.    \\
 {\bf 2)}Limit of  an ultrashort pump pulse $ { \taup  \ll |\taus| \le |\tauipm| }$.  In the opposite limit, the narrow pump profile forces $\bt_s = \eta \bt_i$ inside the box function, and
\begin{align}
\phi(\bt_s,\bt_i) \underset{ \taup \ll |\taus|} {\longrightarrow}  &
\frac{g}{2|\tauispm|}
\alpha_p \left(\frac{\bar{t}_s-\eta\bar{t}_i}{1-\eta}\right)
{\rm Rect}\left[\frac{\bar{t}_s}{\taus} \right] 
\label{asym2}
\end{align}
 Eq.\eqref{asym2} approximately describes the three lower plots of  Fig.\ref{fig4_temp}, where the correlation function is peaked along the line $\bar t_s = \eta \bar t_i$.  In the example (ii) where $\eta=0$, this line is parallel to the $\bti$ axis, and the  joint temporal amplitude is approximately factorable (see Fig.\ref{fig4_temp}f),  implying that the two-photon state  \eqref{statec2} is separable. In  the other two examples with $\eta \ne 0$, the state is entangled. Notice the 
 opposite signs of $\eta$ in  panels (c) and (j) of  Fig.\ref{fig4_temp}, and  that in terms of the original time arguments, by using Eq.\eqref{tA}: $ \bar t_s =\eta \bti $ implies $ (t_s- t_{Ap} )= \eta (t_i -  t_{Ap})$. Thus, in the ultrashort pump regime,    the exit times of twin photons are correlated for $\eta>0$ , while they are  {\em anticorrelated} for $\eta <0$, as a result of the fact that one photon is slower than the pump while the other is faster. \\
 {\bf 3)}{Intermediate  pump pulses} $ {|\tauipm|  \gg \taup  \gg|\taus| }$. Notice that this limit is well defined in the counterpropagating case, where the two scales are naturally  separated, but needs not to exist  in the co-propagating case, where it basically requires that one of the two time constant vanishes, e.g.  $\taus \to 0$ as in the example (ii). If this limit exists, then 
\begin{align} 
\phi(\bt_s,\bt_i)
\underset { |\tauipm|\gg\taup\gg |\taus| } {\longrightarrow}
\frac{g}{2|\tauipm-\taus|} {\alpha}_p\left(\bar{t}_s\right)&
{\rm Rect}\left(\frac{\bar{t}_i}{|\taui-\taus|}\right) \, 
\label{asym3}
\end{align}
 and the  biphoton correlation becomes separable in its arguments,  as approximately shown in  Fig.\ref{fig4_temp}(b),(f), implying that  two-photon state \eqref{statec2} is separable.
\par 
As a final for this section, we notice that despite the fact that the joint temporal probability  as a function of  $(t_i,t_s)$ changes  completely  its shape in different pump regimes, when considered only as a function  of the time difference $t_i-t_s$, it always retains a box-function shape.  This means that if the coincidence count rate \eqref{G2} at the crystal exit faces 
is registered only as a function of the delay between the exit times of the twins, no information is gained about the entanglement or separability of the state. Indeed, if we  rewrite the  temporal correlation in Eq. \eqref{psitemp} as a function of $\delta t = t_i-t_s$, we have 
\begin{align} 
\phi (t_s,t_s+ \delta t ) & \propto \alpha_p \left(\bts- \frac{\eta}{1-\eta} \delta \bar t \right) 
\mathrm{Rect}\left(\frac{\delta \bar t }{\Dtau}\right)  \nn \\
\phi(t_i -\delta t ,t_i) & \propto \alpha_p \left(\bti+ \frac{1}{1-\eta} \delta \bar t \right) 
\mathrm{Rect}\left(\frac{\delta \bar t }{\Dtau}\right)  \nn 
\end{align}
 Then, if  coincidence counts are measured only as a function of the time delay $t_i -t_s$ between twin photons, without detecting the absolute arrival time of the signal or of the idler, 
\begin{align} 
\bar G_{si}^{(2)} ( \delta t ) & = \int dt_s | \phi (t_s, t_s + \delta t ) |^2 = \int dt_i | \phi (t_i -\delta t , t_i) |^2 \nn  \\
&=\frac{ {\cal N} }{2 \Dtau} \mathrm{Rect}\left(\frac{\delta \bar t }{\Dtau}\right) .
\end{align}
where ${\cal N} =g^2   \frac{\sqrt{\pi}\taup} {2 \Dtau}  $ is the total  number  of photon pairs (see Sec.\ref{sec:spectrum}). 
Therefore the coincidence counts as a function of $t_i-t_s$ always reproduces the  flat distribution of time delays characteristic of the spontaneous process, regardless whether the state is temporally entangled or separable. 
\par 
Section \ref{sec:interpretation} will offer an interpretation of these different behaviours of the temporal correlation of twin photons.
Before that,  we make first  a quantitative analysis of the degree of the entanglement of the state. 
\section{Entanglement quantification}
\label{sec:Schmidt}
The degree of entanglement of the state is here characterized by  the  Schmidt number \cite{ekert1995,parker2000}, which also estimates  the number of independent modes participating 
to the entangled state \cite{exter2006}. 
It is defined as the inverse of the purity of the state of each separate subsystem
\beq
\cappa= \frac{1}{ {\mathrm Tr} \{\hat \rho_s^2\} }= \frac{1}{ {\mathrm Tr} \{\hat \rho_i^2\} }
\eeq
where $\hat \rho_s$,  $\hat \rho_i$  are the reduced density matrix of the signal  and  idler
when the PDC state \eqref{state} is conditioned to a photon count.  For example for the signal 
$\hat \rho_s=\frac{1}{  \langle \phi _{2} |\phi _{2}\rangle 
     } 
{\mathrm Tr}_i \{  |\phi _2\rangle \, \langle 
\phi _2|  \} $.  The inverse of the Schmidt number thus  straighforwardly  gives the degree of purity of the heralded photon state, via a single parameter that can be calculated without resorting to the esplicit Schmidt decomposition.  Indeed, 
for a biphoton state of the form \eqref{statec}, 
the Schmidt number can be expressed through an integral formula  \cite{gatti2012,mikhailova2008}
\beq
\cappa = \frac{ {\cal N}^2} {B}   
\label{kintegral}
\eeq
where
\begin{align}
 {\cal N}  = & 
\int d\Oms   \int d\Omi \,  \left| \psi(\Oms, \Omi)   \right|^2  
\label{enne} 
\end{align}
is the mean number of signal or idler photons $\langle  \hat  N_s \rangle = \langle  \hat  N_i \rangle$ generated by the pump pulse; 
\begin{align}
B= & \int d\Oms \int  d\Oms' \int  d\Omi \int d\Omi '    \left[    \psi(\Oms, \Omi )    \psi(\Oms', \Omi ')    \right.  
\left. \psi^* (\Om_s,\Om_i')  \psi^* (\Om_s',\Om_i)  \right]   \, , 
\label{B} 
\end{align}
represents their normally ordered fluctuations(see \cite{gatti2015}):  $B=\langle: \delta\hat  N_s^2 : \rangle=\langle: \delta\hat  N_i^2 : \rangle $.
\par
\begin{figure} 
\includegraphics[scale=0.69]{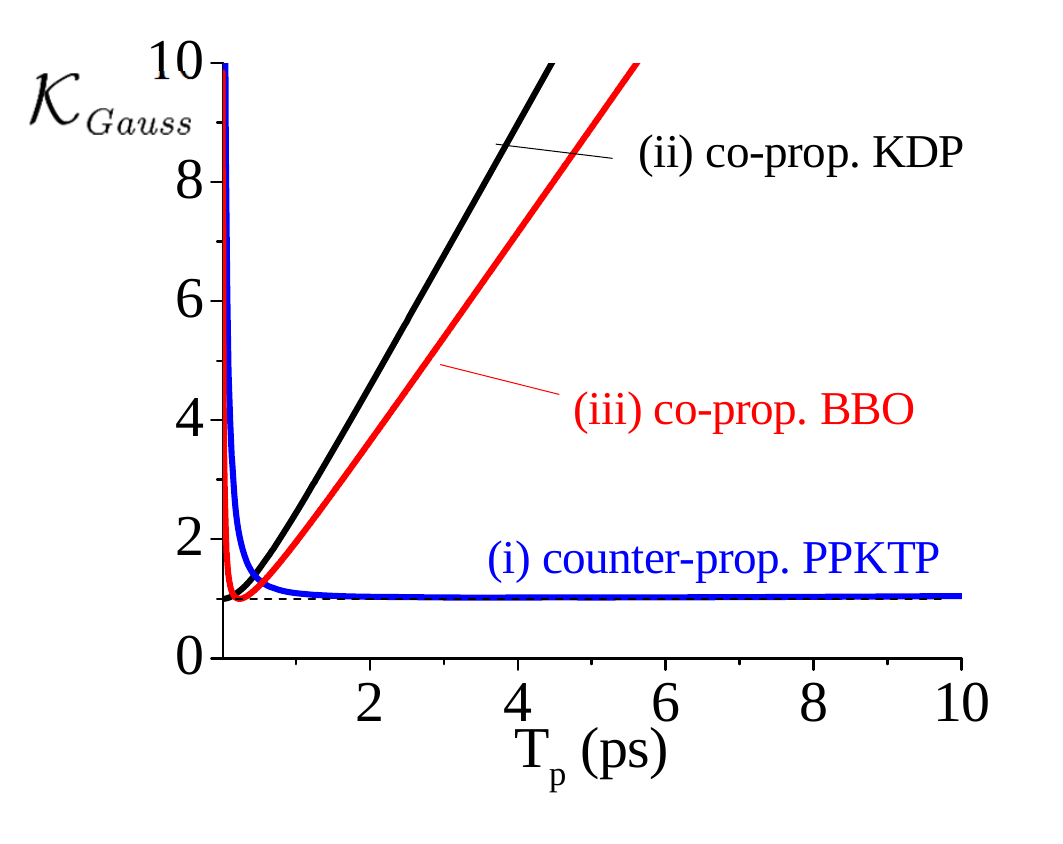}
\caption{ Schmidt number $\cappa$    in  the Gaussian approximation [Eq.(\ref{Kgauss})],   
as a function of the pump duration for   (i)  counter-propagating PPKTP,  (ii) co-propagating KDP $\eta=0$,  (iii) co-propagating BBO  $\eta=-1$(parameters  in Table \ref{table1}).   In  case (i)  separability is achieved  in a much broader  range of pump durations and for longer pulses. }
\label{figKtaup}
\end{figure}

In this work the Schmidt number $\cappa$ is calculated in two ways: (i) "exact " results are  obtained  by numerical integration 
of  Eqs.(\ref{kintegral})-(\ref{B}),  where   the complete Sellmeier dispersion formula in \cite{niko91,kato2002,zernike64} are used; (ii) approximated analytical results are derived within the "Gaussian approximation"  of  the spectral amplitude extensively used in  former  studies of   co-propagating PDC \cite{grice2001,uren2006}. This  consists in  replacing  the sinc function in Eq.(\ref{psi}) by  a Gaussian of its argument, and 
then  using the linear approximations \eqref{Dlin} and  \eqref{betalin},  obtaining thereby
\begin{align}
\sinc\frac{\DDpm(\Oms,\Omi)l_c}{2}&\to e^{-\gamgauss\left[\frac{\DDpm(\Oms,\Omi)l_c}{2}\right]^2}
\approx  e^{ -\gamma \left(\taus\Oms+\tauipm\Omi\right)^2}
\label{sinc_apr}
\end{align}
where e.g. $\gamgauss=1/6$  if one equates the leading order of the  Taylor expansions of the $\sinc$ and the Gaussian, or $\gamma=0.193$ if one 
requires that they have the same full width at half maximum.     
In addition, one has to consider a Gaussian pump pulse 
$\alpha_p(t)= e^{-t^2/2\taup^2}$,  of duration $\taup$ and spectral width $\Delta \Om_p=1/\taup$, so that 
$
\tilde{\alpha}_p(\Om)=\taup e^{-\frac{\taup^2}{2} \Om_p^2},
$
The  spectral  amplitude (\ref{psi})  takes then the Gaussian  form : 
\beq
\label{psi_gauss}
\psi^{\sppm}(\Oms,\Omi)\to
\frac{g\taup}{\sqrt{2\pi}}
e^{i\left[t_{As} \Oms+  t_{Ai}   \Omi\right]}
e^{-\sum_{i,j=s,i} c_{ij} \Om_i \Om_j}
\eeq
where  constant phase terms have been omitted and the real coefficients $c_{ij}$  [see also \cite{uren2006}] are
\beqa
c_{ss}&=&\frac{\taup^2}{2}+\gamgauss\taus^2,\label{c11}\\
c_{ii}&=&\frac{\taup^2}{2}+\gamgauss\tauipm^2,\label{c22}\\
c_{is}&=&\frac{\taup^2}{2}+\gamgauss\taus\tauipm.\label{c12}
\eeqa 
The approximation \eqref{psi_gauss} allows to extract analytical formula,  providing  a straightforward comparison between the two cases.
Inserting  the formula  \eqref{psi_gauss} in the expression  of  $\cappa$ in  Eqs.\eqref{kintegral}-\eqref{B}, and performing the Gaussian integrals involved, we find a single formula that holds for any PDC configuration: 
\begin{align}
\cappaG &=\sqrt{\frac{c_{ss}c_{ii}}{c_{ss}c_{ii}-c_{si}^2}}
= \left[
1+\left(    1 +   \frac   {  2\gamma\, \taus \tauipm }   {\taup^2} \right)^2 \frac{\taup^2}{2\gamma |\tauipm -\taus|^2}
\right]^{1/2}  \label{Kgauss} 
\end{align}
and gives a simple way of estimating the degree of purity ${1} /{\cappa}$ of heralded single photons for any crystal length and dispersion relations in the medium. We notice that in this formula the various parameters  that  characterize the  PDC process (crystal length, dispersion relations for the three waves)
have been condensed in only two parameters, i.e. in the two time constants $\taus$ and $\tauipm$ that rule  the relative propagation of the three waves in a crystal of length $l_c$. \\
The Gaussian formula \eqref{Kgauss}, seen as a function of the pump duration $\taup$ (see Fig.\ref{figKtaup}),  presents   for long pump pulses a linear asymptote: 
\beq
 \cappaG \to \frac{\taup}{\sqrt{2\gamma} |\tauipm -\taus|} \qquad \text{ for  } \taup \gg \sqrt{ 2\gamma |\tauipm \taus|}. 
\label{KGauss_long}
\eeq
corresponding to an entangled state with $\cappa \gg 1$; 
by shortening the pump  it reaches a minimum, and then it stays close to the curve 
\beq 
 \cappaG \to \frac{ \sqrt{2\gamma} |\taus \tauipm|}{\taup |\tauipm -\taus|}  \qquad \text{ for }\taup \ll \sqrt{ 2\gamma |\tauipm \taus|}
\label{Kgauss_short}
\eeq
$\cappaG$ takes its minimum (the purity takes its maximum) for
\beq
\label{taupmin}
\taup^{min} 
=\sqrt{2\gamgauss|\taus\tauipm|}
\eeq
which is basically the  geometrical mean between the two time constants $|\tau_s|$ and $ |\tauipm|$.
The minimum value of $\cappa$ depends both on the sign and on the magnitude of $\eta=\taus/\tauipm$ and is given by
\beq
\cappaG^{min}
=
\begin{cases}
1&\text{for $  \taus\tauipm \leq 0 \rightarrow  \eta \leq 0$} \\
\frac{\tauipm+\taus^{\sppm}}{\tauipm-\taus^{\sppm}} =\frac{1+\eta^{\sppm}}{1-\eta^{\sppm}} \; &\text{for $\taus\tauipm > 0 \rightarrow  \eta >0 $}
\end{cases}
\label{Kmin}
\eeq
As well known, the Gaussian approximation (\ref{psi_gauss}) predicts
complete separability  $\cappa=1$  for  $\eta \leq 0$, as can be immediately recognized by inspection of the mixed term $c_{is} $ in  Eq. \eqref{c12}. 
The condition  $\eta=0$, corresponds to the asymmetric group  velocity matching of the example (ii), and  the ideal value $\cappa=1$  is reached only asymptotically for  $\taup \to \taup^{min} =0$. The condition  $\eta <0$ requires 
that  $\taus$ and $\tauipm$ have opposite signs (as in the example (II) for $\eta=-1$) and  a separable state can be in principle obtained  for a finite pump  duration  $\taup= \taup^{min}$.   These conditions  straightforwardly lead  to a separable  Gaussian spectral  amplitude \eqref{psi_gauss}. 
\\
Alternatively, for positive $\eta$, the two-photon state can be made almost separable by choosing a
configuration for which $\eta$ is sufficiently small, because 
\beq
\cappaG^{min} = \frac{1+\eta^{\sppm}}{1-\eta^{\sppm}}  \simeq 1+ 2\eta  \qquad \text{for $ 0<\eta \ll 1$}
\eeq
  Notice that this last condition is naturally fulfilled in the counter-propagating case. 
\begin{figure}  
\centering
\includegraphics[scale=0.6]{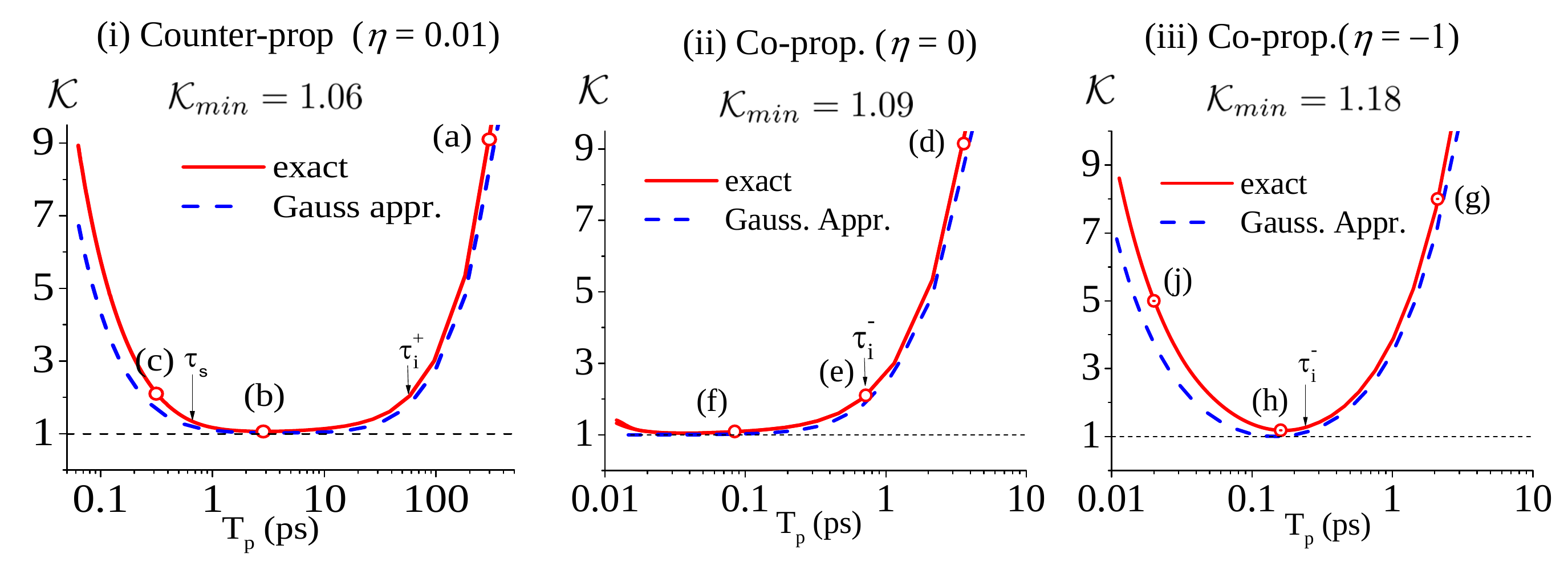}
\caption{Schmidt number $\cappa$: comparison of the Gaussian result (dashed blue lines) and the exact one (solid red lines) 
(i) counter-propagating PPKTP,   (ii) co-propagating  KDP   (iii) co-propagating BBO (parameters  in Table \ref{table1}).  
In (i) the  state is nearly separable for 
$\taup$  intermediate between $\taus=0.67$ps and $\taui=63$ps. 
In (ii) and (iii), 
separability is achieved only for subpicosecond pulses with $\taup\ll\tauib$.   The inverse of $\cappa_{min}$  is the achievable degree of purity.  
The hollow red dots correspond to the plots   in Figs.\ref{fig4_temp} and \ref{fig4} . 
}
\label{fig3}
\end{figure}
\par 
Figures \ref{figKtaup} and \ref{fig3} plot the Schmidt number $\cappa$  as a function of 
the pump duration, for  the three examples described in Table \ref{table1}. 
Fig.\ref{figKtaup} superimpose the three results,  calculated  in the Gaussian approximation  (\ref{Kgauss}), while 
Fig.\ref{fig3} (notice here the horizontal logarithmic scale) shows,   for each example, a comparison between   the Gaussian result  and the exact one, obtained by  numerical integration of  Eqs.(\ref{kintegral})-(\ref{B}).  The  Schmidt number curves reproduce in a more quantitative way  the behaviour of the temporal correlation analysed in Sec. \ref{sec:temporal} (notice that the  hollow dots in Fig.\ref{fig3} correspond to the parameters of the plots of in Fig.\ref{fig4_temp}): for long pump pulses the state is entangled in all the examples,  by shortening the pump pulse $\cappa$   reaches a minimum close to the ideal value $\cappa^{min}=1$ and then  it grows again in the two examples with $\eta \ne 0$, while it   stays close to the minimum in the example (ii) with $\eta=0$.
However, 
 evident from these figures are  the different ranges of pump durations  $\taup$ in which 
 separability can be achieved: the counter-propagating case  displays  a broad plateau with $\cappa \simeq 1$ for  $\taup$ in the range   between $2$ps   and $10$ps, while  in
the  co-propagating case separability requires supbicosecond pulses, of duration 
$\taup\lesssim 100$fs $\ll\tauib=720$fs in the $\eta=0$  case  (ii), and 
$\taup\simeq \taup^{min}=147$fs in the $\eta=-1$  case (iii). \\
Fig. \ref{fig3} also shows  some 
discrepancy between the  Gaussian  results and the exact ones, especially for short pump pulses. In particular, the minimum  of $\cappa$ is always slightly larger than the Gaussian result \eqref{Kmin},  and  never reaches the ideal value $\cappa=1$  even for $\eta \le 0$.
Clearly these discrepancies have to be ascribed both to the replacement of the sinc by a Gaussian and to the effect of  dispersion, which becomes relevant only for short pump pulses, and/or long crystal.  Because of this, 
the degree of purity achievable in the counter-propagating configuration  with $\eta=0.01$  is comparable  to what  obtained  in the two cases with $\eta\le0 $,   which require much shorter pump pulses and stringent phase matching conditions. 
\par 
It is also interesting to see how these results reflect in the shape 
\begin{figure*} [ht]
\centering
\includegraphics[scale=0.65]{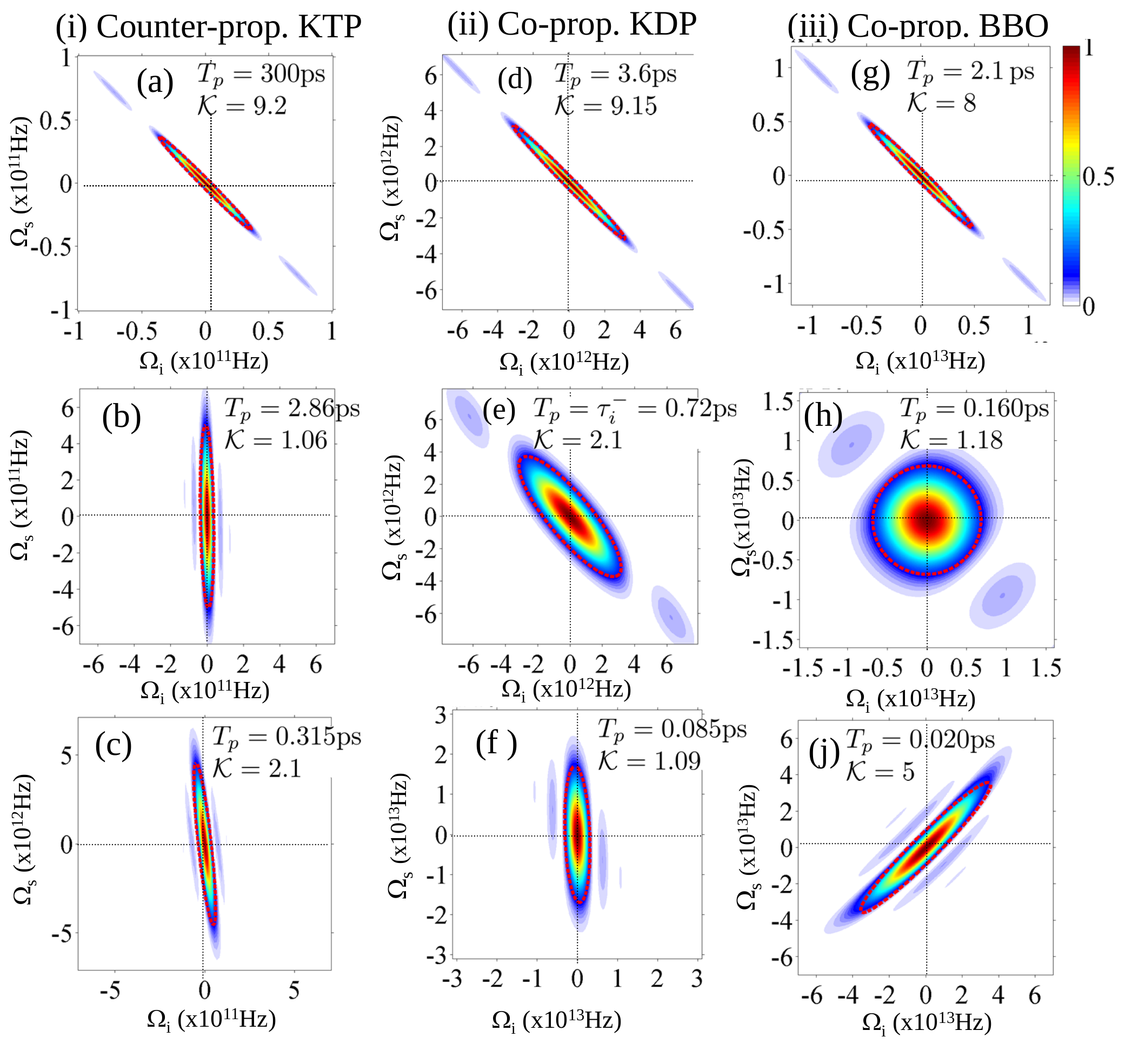}
\caption{Joint spectral  probability $|\psi(\Om_s,\Om_i)|^2$ plotted for pump pulse durations   decreasing from top to bottom, for the same parameters in in Fig.\ref{fig4_temp} and in correspondence of  the hollow red dots  in Fig.\ref{fig3}.  Panels (b), (f) and (h) represents  nearly separable situations. The red ellipses  are the  isoprobability curves of the Gaussian formula \eqref{psi_gauss}. 
Notice that in panel (c) the idler  scale $\Omega_i$ is zoomed by a factor 10 with respect to the signal 
scale.}
\label{fig4}
\end{figure*}
of the joint spectral probability $|\psi|^2$  in the $(\Om_i, \Oms)$  plane,  plotted in Fig.\ref{fig4} for the same parameters as in Fig. \ref{fig4_temp}.   The red ellipses  here represent the curves
$
c_{11}\Om_s^2+c_{22}\Om_i^2+2c_{12}\Om_s\Om_i= 1
$
where according to the Gaussian formula (\ref{psi_gauss}) $|\psi|^2$ reduces by $1/e^2$, and permit to estimate visually the validity of the Gaussian approximation. 
The spectral correlation in this figure  gives the complementary view   with respect to  the temporal correlation in Fig.\ref{fig4_temp}, but obviously displays the same amount of entanglement: 
For long pump pulses  (top raw) the state is highly entangled in all the three examples, with the biphoton amplitude peaked along the diagonal $\Oms=-\Omi$ where energy conservation takes place.  For short enough pump pulses [bottom panels (c) and (i)] the state is again entangled, with the biphoton amplitude peaked along the line 
$\Oms=-\eta \Omi$ where momentum conservation, i.e. phase matching, is realized.  Panels (b) (f) and (h) correspond to the optimal conditions for separability, and we see that in these cases the biphoton probability is a nearly  factorable function of its arguments, being approximately an ellipse with its principal axes aligned along  $\Omega_s$ and $\Omega_i $. 
\section{Interpretation and discussion}
\label{sec:interpretation}
In order to understand these behaviors, and to be able to describe the  peculiarity of the co-propagating  geometry (case $\eta \approx -1$ ), we resort to the very notion of temporal correlation between the members of a pair, which relies on the possibility of predicting the arrival time of one photon by detecting  the arrival time of its twin, with a precision better than what would be obtained with an unconditional measurement. \\
Referring for definiteness to the signal photon (the one that  in any case co-propagates with the pump), we first wonder what is the uncertainty in its arrival time conditioned to  detection of the idler.   Then we will compare it with the uncertainty in its arrival time not conditioned on the detection of the idler. 
\\
There are actually two distinct mechanisms 
 by which the  exit time  of one photon can be ascertained by detection its twin:  one is simply based on their simultaneity,  the other  relies on the possibility to gain information on the point where down-conversion took place. \\
The first one prevails for { long pump pulses} $ { \taup  \gg |\tauipm|\ge |\taus|}$. When  the duration of the pump pulse is much longer than the delays of twin photons with respect to its center, then
 the idler and signal  wavepackets propagate under the pump pulse, and  detection of neither photon gives  information on the point of the crystal where the pair was generated. Thus, our ability  to infer the arrival  time of one photon by detection of its twin is limited by the degree to which they arrive simultaneously at their end faces,  i.e. by  the spread of the distribution of  their mutual time delays. This  is described by 
the box function in Eqs. \eqref{psitemp} and\eqref{asym1},   characteristic of  the spontaneous  PDC processes, where  a  photon pair  can be generated at any point along the crystal with uniform probability.  Then,  depending on the point where down-conversion took place, the delay between the exit   of the twins has a flat distribution  in the interval $[- \Dtau, + \Dtau ]$ \footnote{For the original time arguments, 
$t_s-t_i = \bar t_s -\bar t_i + (t_{As} -t_{Ai} )$. In the co-propagating case  the box function  is then shifted to the interval $[0,2\Dtau]$ while in the copropagating case the shift is negligible.  }
where  
$
\Dtau:=\frac{l_c}{2}\left|\frac{1}{v_{gs}}\pm\frac{1}{v_{gi}}\right|
$.
Notice that counter-propagating  photons  ($+$ sign)  can be  delayed up to their transit time along the sample, since they appear on opposite faces of the crystal. Conversely,  co-propagating twins ($-$ sign) exit from the same face,  and appear with a small delay  ruled by their GVM, the extreme case being  when they propagate exactly at the same velocity, and  always exit  simultaneously, in which case the width of the box function vanishes. According to this mechanism  the exit time of the signal can  be deduced from  that of the idler as 
\begin{align}
\quad \bar t_s = \bar t_i &\qquad \text{within} \; \tcorr \simeq \Dtau= | \tauipm -\taus |&\qquad \text{for }\;
  \taup  \gg|\tauipm| 
\label{tcorr_long} 
\end{align}
where $\tcorr$ indicates the spread of the distribution of the signal arrival time conditioned to detection of the idler, i.e the {\em correlation time}.
\par 
The second mechanism, described  by the factor $\alpha_p \left(\frac{\bar{t}_s-\eta\bar{t}_i}{1-\eta}\right)$ in Eq. \eqref{psitemp},  prevails for  ultrashort pump pulses $ \taup \ll |\taus| \le |\tauipm|$. In this case  the twin photons  tend to separate from the pump  during propagation, and detection of any of them
 provides an indication of the point where the pair was generated.  By using a rough picture of photons as wavepackets propagating without deformation  with their group velocities, one can imagine that if a photon pair was generated at a point $z_0$, and conversion occurred from a pump photon delayed by  $ \delta t_p $ from the centre of the pump pulse, than the idler photon will arrive at its end face at time 
\beq
t_i  =\begin{cases} \frac{z_0 }{v_{gp} } + \frac{z_0 }{v_{gi}}  + \delta t_p \quad  &\text{counter-prop. case } \\
\frac{\ z_0 }{v_{gp} }  + \frac{ l_c-z_0}{v_{gi}}  + \delta t_p  \quad  &\text{co-prop. case } 
\end{cases}  
\eeq
i.e, for  the  barred time argument $\bar t_i =   t_i - \frac{1}{2}\left( \frac{l_c}{v_{gp} } + \frac{l_c}{v_{gi} } \right) $: 
\begin{align}
\bar t_i &=   
\left(z_0 - \frac{l_c}{2} \right) \frac{2}{l_c} \tauipm + \delta t_p 
\label{ti}
\end{align} 
The same argument gives the arrival time of its twin signal photon as
\begin{align}
\bar t_s  
&= \left(z_0 - \frac{l_c}{2} \right) \frac{2}{l_c} \taus + \delta t_p
\label{ts}
\end{align} 
where in the above formulas  $ \delta t_p $ can be considered as a Gaussian stochastic variable with variance $\taup/\sqrt{2}$, because a pump photon can be downconverted at any time along the pump pulse, with a probability proportional to the Gaussian pump intensity. 
Then, by comparing Eqs (\ref{ti}) and (\ref{ts}): 
\begin{align}
\bar t_s  = \bar t_i  \frac{\taus }{\tauipm} + \delta t_p \left(  1- \frac{\taus }{\tauipm} \right) 
= \eta \bar t_i + (1-\eta) \delta t_p 
\end{align} 
 Thus, according to these arguments, the arrival time of the signal conditioned to detection of the idler  is
\beq
\quad \bar t_s = \eta \bar t_i  \quad \text{within} \; \tcorr
 \simeq T_p (1-\eta)  \quad ( \taup \ll |\taus| )
\eeq
\par
These conditional uncertainties have to be compared with  the unconditional uncertainty in the arrival time of the signal., i.e.  the  spread of the distribution of  the exit  times of the signal when the idler is not detected.  Also here there are two sources of uncertainty, one is the width $\taup$ of the  Gaussian distribution of the pump, because the photon can be downconverted from any portion of the pump pulse, the other one is the point of the crystal where down-conversion took place. Depending on the latter, the delay of the signal photon with respect to the center of the  pump pulse ranges in the interval [0, $2 |\taus|]$ with uniform probability. Clearly,  the first mechanism dominates for $\taup \gg |\taus|$,  as shown by  Eqs. \eqref{asym1} and \eqref{asym3}, the second mechanism for  $\taup \ll |\taus|$ [see Eq. \eqref{asym2}]. A more precise analysis, based  on the temporal correlation\eqref{psitemp}  gives

\begin{align}
(\delta t_s )_{\mathrm uncond.} &= \sqrt{\frac{T_p^2}{2} + \frac{\taus^2}{3}  }\to 
\begin{cases}
\frac{T_p}{\sqrt{2}} &  \taup \gg |\tau_s|\vspace{4pt } \\
\frac{|\tau_s|}{\sqrt{3}} &  \taup \ll |\tau_s| 
\end{cases}
\label{unconditional} 
\end{align}
where $(\delta t_s )_{\mathrm uncond.} $ is the variance of the distribution of the arrival times of the signal when the idler is not detected. 
\par
Summarizing the results, we have the following situation: 
\par 
\noindent 
{\bf 1) For a long pump pulse} $ \boldmath{ \taup  \gg |\tauipm|\ge |\taus|}$, comparing  the unconditional and  conditional  uncertainties 
 in the arrival time of the signal, in  Eqs. \eqref{unconditional}  and 
\eqref{tcorr_long}, 
we see that    for such long pumps $\taup \gg |\tauipm|$,  
the state is in general  highly entangled  because \footnote{Notice that  $ 0<(1-\eta) \le 2$,  because of  Eq.\eqref{etarange}}. 
\beq
\frac{ (\delta t_s )_{\mathrm uncond.}}  { \tcorr}
\simeq \frac{\taup}{ |\tauipm -\taus| } = \frac{\taup}{ |\tauipm|  (1-\eta) }\gg 1
\label{ratio1}
\eeq
Euristically, the above ratio  also gives an estimate of the number of entangled modes, 
well in accordance with the asymptotic behaviour  of the Schmidt number in Eq.\eqref{KGauss_long} for long pump pulses.\\
\noindent
{\bf 2) For an ultrashort pump pulse}  $\boldmath \taup \ll |\taus| \le |\tauipm|$,  
the state is again  highly entangled  because 
\beq
\frac{ (\delta t_s )_{\mathrm uncond.}}  { \tcorr}
 \simeq  \frac{|\taus|}{ \taup(1-\eta) } = \frac{|\taus \taui|}{ \taup|\tauipm -\taus| }  \gg 1
\label{ratio2}
\eeq
for $\taup \ll |\taus|$.  Also in this case the ratio in Eq.\eqref{ratio2} can be considered an estimate of  the number of entangled modes, and indeed reproduces the asymptotic behaviour of the Schmidt number in Eq.\eqref{Kgauss_short}. 
\par
Following the above arguments,  in particular the results \eqref{ratio1} and  \eqref{ratio2}, there is no chance that the state become separable when $\taus \simeq \tauipm$,  i.e. $\Dtau\simeq 0$. When the  two photons propagate in the same direction at similar velocities,  in fact,  detection of one photon will always provide an extremely precise  information about the arrival time of the other, basically  because they exit the crystal  almost simultaneously.  
This is the typical situation that occurs in the co-propagating case, in the absence of any velocity matching strategy, and explains why co-propagating photons in general display high temporal entanglement. 
\\
Conversely, in  the counterpropagating geometry  the exit times of the twins lack simultaneity
because they propagate in opposite directions, and this case is naturally characterized by a strong asymmetry between 
the twins, in particular  they have very different delays   $\taui \gg |\taus|$ from the pump.  Because of that,  the separable limit \eqref{asym3} can be always realized  for intermediate pump durations  $ \taui\gg \taup \gg |\taus| $ which need not to be ultrashort because $\taui$ is a long time scale.  This limit physically corresponds to a situation where the forward  photon (here the signal) always propagate below the pump pulse, but the pump pulse   is short enough that the backward  photon rapidly separate from it.  In these conditions the temporal localization of the pump provides an absolute timing information on the exit time of the signal,  as precise ($\sim \taup$) as  the information that can be gained by detecting the idler $\sim \taup (1-\eta) \approx \taup$, and  the exit times of the twins appear uncorrelated. An other way of looking at the situation is that the forward photon is locked to the pump so that it cannot provide any information on the point where down-conversion occurred,  and by detecting the signal one does not gain any more precise information on the exit time of  the backward idler than by not detecting  it.  
\par
In the co-propagating configuration, the same limit can be  reached only by creating a strong asymmetry between the propagation velocities of the two photons relative to the pump  $|\taus| \ll |\tauib| $, which in practice requires that  the signal  is velocity matched to the pump  \cite{grice2001, mosley2008b, mosley2008}  as in the example (ii).  However, in order that the idler separate from the pump pulse along a finite  propagation length, the  pump pulse must be  in this  case ultrashort (Fig.\ref{fig4_temp}f). As a consequence, as we shall see in the following, the generated twin photons  are  broadband. 
\par
An alternative strategy,  which can be realized only in the co-propagating case,   is the so-called  symmetric velocity matching \cite{grice2001, zhang2012}  where $\taus =-\tauib$ and $\eta =-1$, as in the example (iii).  This condition, which is rather difficult to achieve,  requires that one photon propagate faster than the pump while the other is slower, so that they are symmetrically delayed with respect to the pump center. 
Negative  values  of $\eta $ are in  general favorable to separability,   because  in this case   $\Dtau =|\tauipm | + |\taus| >|\tauipm |, |\taus|  $, and  twin photons are more simultaneous with the pump than between themselves. 
 In these conditions, a properly localized  pump pulse may provide a  timing of the  exits of the twins  more precise than the information that can be inferred from detection of any of them.  Focusing e.g  on the case $\eta$=-1, when the pump duration 
  $\taup$ approaches  $|\tauib| =|\taus|$, 
 both photons are still locked under the pump, so they do not provide sufficient information on the point where generation occurred, and on the other side they  are no more simultaneous than the pump duration  $\Dtau =2|\tauib| >\taup$. 
\footnote{ Quantitatively, 
when the pump duration approaches $|\tauipm| $ in Eq. \eqref{ratio1}, or $|\taus| $ in Eq. \eqref{ratio1},  then the ratio 
$\frac{ (\delta t_s )_{\mathrm uncond.}}  { \left. (\delta t_s )\right|_{\bar t_i}}
\to 
 \frac{1}{  1+ |\eta| } <1 $.} 
\\
This configuration produces rather  pure  photons (see Figs.\ref{fig3}(iii) and \ref{fig4_temp}h). However, their purity is rather far from the  ideal result $\cappa_{min} =1$  predicted 
by  the Gaussian formula  \eqref{Kmin},  
which seems  rather an artifact of replacing the sinc by a Gaussian. 
Actually, if  one looks at  Fig.\ref{fig4_temp}h, which  corresponds to the minimum  of $\cappa$   for $\eta=-1$,  the joint temporal probability exhibits a diamond-like shape,  which does not look really  separable in its arguments, and indeed $\cappa_{min}=1.18 $  is  slightly  higher  than in  the other two cases. Notice that in contrast the biphoton probability in the spectral domain in Fig.\ref{fig4}h has a nearly circular shape which looks much more factorable (actually in this case deviations from factorability are indicated by the sidelobes of the sinc functions
) 
 This demonstrates the usefulness of taking the two complementary views in the  temporal and spectral domains. 
Such imperfect separability  is a consequence of the sharp boundaries of the nonlinear medium,  which are at the origin of  the  rectangular box function appearing in  Eq.(\ref{psitemp}).  Taking  the Gaussian approximation (\ref{psi_gauss}),   this box function   is replaced by a smooth 
 Gaussian of the same width,   ${\rm Rect}\left(\frac{\bar{t}_s-\bar{t}_i}{\Delta \tau}\right) \to  \exp\left(-\frac{(\bt_s-\bt_i)^2}{4\gamma \Dtau^2}\right)$,  and then the temporal correlation   becomes factorable: 
\begin{align}
\phi(\bt_s,\bt_i)
&\propto
{\alpha}_p\left(\frac{\bt_s+\bt_i}{2}\right)  
{\rm Rect}\left(\frac{\bar{t}_s-\bar{t}_i}{\Dtau}\right)  
\to
e^{-\frac{\bt_s^2+\bt_i^2}{4\gamma(\Dtau)^2}}   
&  \text{for $\taup= \taup^{min}= \sqrt{2\gamma} |\tauib|$}
\label{phi_fact}
\end{align}
(the same result is obtained by directly Fourier transforming the Gaussian approximation of $\psi(\Om_s,\Om_i)$
in  Eq.(\ref{psi_gauss})).
 Notice that the sharp crystal  boundaries have less impact on the separability when  $|\eta| \ll 1$, as in the examples i) and ii), because of the elongated shape of the temporal correlation (see   Fig.\ref{fig4_temp}b and f).  These effects might be eliminated and  the purity of the heralded photons increased by  engineering  a Gaussian  nonlinearity profile of a poled crystal, as demonstrated by Branczyk et al. \cite{branczyk2011}.

\section{Spectro-temporal properties of heralded photons}
\label{sec:spectrum} 
Once  assessed the conditions under which pure single heralded photons can be obtained,  
it is clearly important to study their individual properties  in view of their use in quantum communication/metrology protocols (coupling with atoms, interferometry etc.).
\par 
For spontaneous PDC, the marginal statistics of individual photons is described by their first order coherence functions. In the temporal domain, for example, these functions are: 
\beqa
G_s^{(1)}(t_s, t_s') &=&\langle \hat a_s^{\dagger}  (t_s) \hat a_s (t_s') \rangle 
= \int d t_i  \phi^*(t_s, t_i)\phi(t_s', t_i)
 \label{G1s}\\
G_i^{(1)}(t_i, t_i') &=&\langle \hat a_i^{\dagger}  (t_i) \hat a_i (t_i') \rangle 
= \int d t_s  \phi^*(t_s, t_i)\phi(t_s, t_i') 
\label{G1i}
\eeqa
with analogous definitions in the spectral domain. 
 In terms of its temporal  $ G^{(1)}$,  the  reduced state of the heralded photon takes the form 
\begin{align}
\hat \rho_j &= \frac{1}{\cal N} \int dt \int dt' G_j^{(1)}(t', t) \, \hat a_j^\dagger (t) |0 \rangle\, \langle 0| \hat a_j (t')
\nn \\
&=\frac{1}{\cal N} \int dt \int dt' G_j^{(1)}(t', t) \, |t \rangle_j \phantom{i}_j\langle t'| 
\label{rhoj}
\end{align}
which in general represents a mixed state, unless the coherence function  $ G_j^{(1)}(t, t')$ is factorable in its arguments. \\ 
For equal times, $G_j^{(1)}(t, t) = \langle \hat A_j^{\dagger\, \mathrm{out}}  (t) \hat A_j^{\mathrm{out}}   (t) \rangle = {\cal I}_j(t) $ gives the temporal intensity  profile of the j-th wave, i.e. the probabilty distribution  of detecting the photon at time $t$ at the crystal exit face.  The spectral distribution of the heralded photon can be instead obtained via 
\begin{align} 
S_j(\Om) = \langle \hat A_j^{\dagger\, \mathrm{out}}  (\Om) \hat A_j^{\mathrm{out}}   (\Om) \rangle= 
\int \frac{dt \, dt'} {2\pi} e^{i\Om (t-t')} G_j^{(1)}(t, t')
\end{align} 
For the sake of brevity  we limit our analysis to the cases of A) long pump pulses, where the state is highly entangled, and B)  conditions where  nearly separable states can be realized. 
\subsection{Long pump pulses, high entanglement}
According to the previous results, when $  {\taup  \gg |\tauipm|\ge |\taus|}$ the two-photon state state is highly entangled in any configuration. We can for example use the asymptotic form \eqref{asym1} of the joint temporal  amplitude\eqref{psitemp}, and insert it into formulas  (\ref{G1s},\ref{G1i}). The coherence functions  of the signal and the idler obtained in this way  are  identical (a part from negligible offsets),  and have the form
\begin{align}
G_s^{(1)}(t, t')&= G_i^{(1)}(t, t') 
  \to \frac{g^2}{2 \Dtau } \left| \alpha_p (t) \right|^2 \mathrm{T} \left( \frac{t'-t}{2 \Dtau  }  \right)
\label{G1long}
\end{align}
where $ \Dtau=|\tauipm-\taus|$, and 
\beq
\mathrm{T} \left( \frac{\delta t }{2\Dtau }  \right) = \begin{cases} 1-\frac{|\delta t |}{2 \Dtau } &
 \delta t \epsilon (-2 \Dtau, +2 \Dtau) \\								
0 & \text{elsewhere} \end{cases} 
\eeq
is the triangular function, which
has the shape of a triangle of base $ 4\Dtau$ and unitary height.  This result is the generalizazion of formula (46)  in Ref.\cite{corti2016}. Actually,  it holds for any PDC configuration provided that the bandwidths in play are not too broad, because it makes only use of the linear approximations \eqref{Dlin} and \eqref{betalin}.   It shows that  when the  pump pulse is much longer than the two characteristic  time scales, the  twin photons have identical properties. The width $\Dtau$  of $G^{(1)} (t,t')$  as a function of the time difference $t'-t$ is  their
{\em  coherence time},   which in this limit is equal to their mutual  correlation time [see Eq. \eqref{tcorr_long}].  
Conversely, the temporal distributions  of twin photons  follow  the profile of the much longer pump pulse: 
\beq
{\cal I}_s(t) ={\cal I}_i(t)  = \frac{g^2}{2 \Dtau  }   \left| \alpha_p (t) \right|^2
\eeq
Notice that from the point of view of classical statistics of light, this  behaviour of the $G^{(1)}$,  with the peak at $t=t'$ much narrower of the intensity distribution is typical of multimode incoherent light. On the other hand, in the quantum description, the state of the heralded photon in Eq. \eqref{rhoj} is in this case   mixed, because  $ G_j^{(1)}(t, t')$ in Eq.\eqref{G1long}  is not  factorable in its arguments. \\
The spectra of the photons have the usual $\sinc^2$ shape characteristic  of spontaneous processes 
\begin{align}
S_s(\Om) = S_i(\Om) = \frac{g^2}{\sqrt{4\pi}} \taup \sinc^2 (\Om \Dtau) , 
\label{S}
\end{align}
and their  spectral bandwidths $\Delta \Omega_j = \frac{1}{\Dtau} $ are the inverse of the correlation-coherence times. Clearly, this explains why counterpropagating twin-photons are  narrowband (order 10 GHz), while co-propagating twin-photons are in general  broadband  (order Thz or more) \\
\begin{figure*} 
\centering
\includegraphics[scale=0.65]{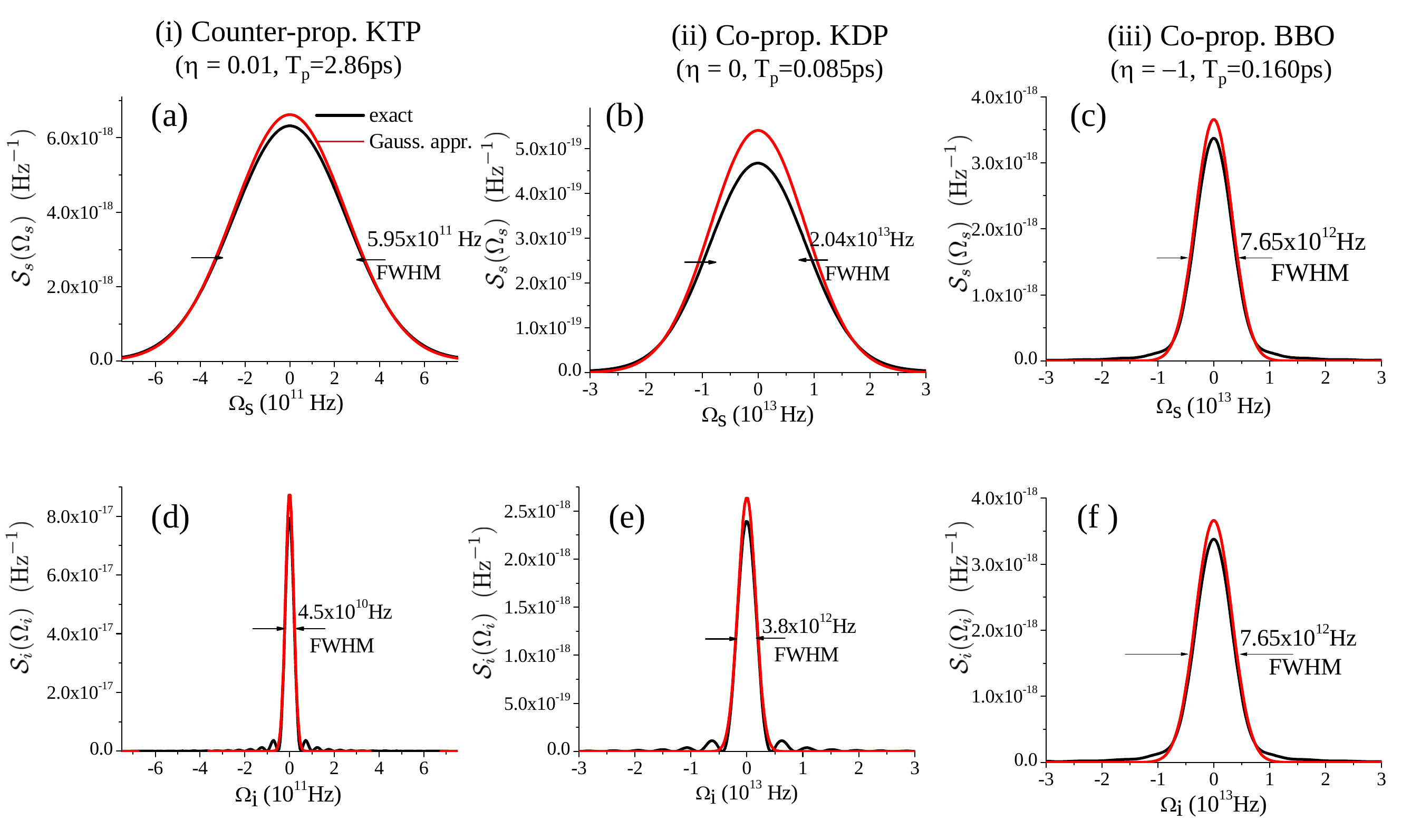}
\caption{Spectra of the signal (top) and of the idler (bottom)  for the three examples considered in Table \ref{table1}, calculated in conditions of separability of the state, 
corresponding to panels (b), (f), and (h) of Fig.\ref{fig4}. Black lines: exact numerical results.  Red lines: Gaussian approximated ones.  Notice in column (i) and (ii)  the Gaussian profiles of the signal and the  narrower  $\sinc^2$ profiles of the idler spectra  as predicted by Eq.\eqref{Ss} and \eqref{Si}.}
\label{fig6}
\end{figure*}
\subsection{Nearly separable regime}
We start from the case  $\eta \ll 1$, including    both the counter-propagating and  the co-propagating case with asymmetric group velocity matching. We remind that for a pump pulse intermediate between the two time scales $|\tauipm| \gg \taup \gg |\taus|$,  one photon propagates locked under the pump  while the other separates from it after being generated. For this reason the  properties of the two photons are  strongly asymmetric, as e.g.  shown by the spectra in column (i) and (ii) of Fig.\ref{fig6}.
  By using the  asymptotic limit\eqref{asym3} of the joint temporal amplitude, and inserting  it into formulas  (\ref{G1s},\ref{G1i}), we obtain for the signal photon: 
\begin{align}
G_s^{(1)}(\bts, \bts') 
&=\frac{g^2}{2 \Dtau}  \alpha_p ^*(\bts)  \alpha_p (\bts') 
\end{align}
which implies that 
\begin{align}
&{\cal I}_s ( t ) = \frac{g^2}{2 \Dtau}  \left|\alpha_p ( t-t_{As} ) \right|^2 ,  \\
&{S}_s (\Om) = \frac{g^2}{2 \Dtau}    |\tilde \alpha_p (\Om) |^2
\label{Ss}
\end{align}
i.e.  the photon that propagates locked under the pump entirely inherits  the spectro-temporal properties  of the coherent pump laser. \\
Conversely for the idler photon 
\begin{align}
G_i^{(1)}(t, t') 
&=  \frac{g^2}{2 \Dtau}  \frac{ \sqrt{\pi} \taup} {2\Dtau}  
\mathrm{Rect }\left( \frac{\bti}{\Dtau}   \right) \mathrm{Rect }\left( \frac{\bti^\prime}{\Dtau}   \right)  , 
\end{align}
and its properties depend on the dispersion properties  of the crystal and on its length, through the parameter 
$\Dtau= |\tauipm-\taus|$. In particular
\begin{align}
&{\cal I}_i (t) =  
\frac{g^2}{2 \Dtau } \frac{ \sqrt{\pi} \taup} {2\Dtau}  
\mathrm{Rect }\left( \frac{t-t_{Ai}}{\Dtau}   \right) , \\
&{S}_i (\Om) = \frac{g^2}{\sqrt{4\pi}} \taup \sinc^2 (\Om \Dtau), 
\label{Si}
\end{align}
The temporal distribution of the idler photon is a rectangular pulse of duration $2 \Dtau$,  because for such a short pump 
the idler photon distribution reflect simply the fact that it  can be generated anywhere in the crystal with uniform probability.  Remarkably,
since $ \Dtau\simeq |\tauipm | \gg \taup$,   the idler  wavepacket  has a longer duration  than the pump itself. In particular, in the counterpropagating case, its duration roughly corresponds to the transit time of the pump along the crystal. 
 The idler spectrum retains the same $\sinc^2$  shape as in the long pump limit, with a bandwidth  $\Delta \Omega_i = 1/ \Dtau \ll \Delta \Omega_p $ .  Hence,  when the state is approximately separable, the heralded idler photon is not only pure, but also  more monochromatic than the pump laser that drives the process.  
 However ,  only  in the case (a) this result  really means that the generated  idler photon is  narrowband, as shown by figure \ref{fig6}d, while   in case b) it is anyway quite  broadband (see Fig.\ref{fig6}e), because separability requires  ultra-broadband pump pulses.
Interestingly,  in the  counterpropagating case  the strong asymmetry between the twin photons, and the fact  that the idler is more monochromatic  than the pump, reflect the unusual coherence properties of the classical signal and idler field generated  above the  MOPO threshold, described in \cite{canalias2011,Montes14}.\\
The  forms of the   coherence functions  of the signal and idler,  $ G^{(1)} (t,t') = \sqrt{ {\cal I} (t)} \sqrt{{\cal I} (t')} $, 
is typical of  single-mode  light, which possess classical temporal coherence: indeed it means that the average length of a temporal fluctuation (the coherence time)  is equal to the duration of the wave-packet, implying that the statistics is a sort of frozen in time.  In the quantum description the states of the heralded signal and idler photon are pure $\hat \rho_j = |\psi_j\rangle\, \langle \psi_j |$, with 
\begin{align} 
|\psi_s \rangle 
&=\frac{1}{(\sqrt{\pi} T_p)^{1/2} } \int dt_s  \alpha_p  \left(t_s-t_{A_s} \right) \, \hat a_s^\dagger (t_s) |0 \rangle \\
|\psi_i \rangle
&= \frac{1}{  \sqrt{2\Dtau}  }  \int dt_i  \mathrm{Rect }\left( \frac{t_i-t_{Ai}}{\Dtau}   \right) \,  \hat a_i^\dagger (t_i) |0 \rangle 
\end{align}
\par 
Finally we consider the alternative technique for reaching separability, i.e. the configuration with $\tau_s= -\tau_i$ ($\eta=-1$), 
in which the two photons co-propagate symmetrically delayed with respect to the pump center. As can be intuitively understood, in this case the properties of the twin photons are completely symmetric.
The mere linear approximation does not give particularly expressive results, so that we have to resort to the stronger Gaussian approximation. According to it, separability is reached at $\taup= \taup^{\rm min}= \sqrt{2\gamma} |\tauib|$, where (calculations not reported here)  the state of each heralded photon  becomes pure, 
$ G_j^{(1)} (t,t') = \sqrt{ {\cal I}_j  (t)} \sqrt{{\cal I}_j (t')} $, with 
the temporal and spectral distributions  of the two waves given by 
\begin{align}
{\cal I}_s(\bts ) &= {\cal I}_i(\bti ) \propto e^{-\frac{(t-t_{Aj})^2}{2\taup^2}} = | \alpha_p \left( \frac{\bt_j}{\sqrt{2}} \right) |^2
\\
S_s(\Om) & =S_i (\Om) \propto | \alpha_p \left( \sqrt{2} \Om \right) |^2
\end{align}
i.e. the two waves have similar properties as the pump, but with a slightly longer duration and a slightly narrower spectrum.  As shown by panel (iii) of Fig.\ref{fig6}, these Gaussian results are rather close to the exact ones. 

\subsection{Mean number of down-converted pairs-Efficiency}
An other important point in view of  applications is the  efficiency by which heralded pure photons can be generated. In the separable regime, where the state is approximately single-mode, this ultimately depends  on the probability that  a photon pair  is generated by  a pump pulse crossing the crystal \footnote{ We are assuming an ideal unit efficiency of detection}
For a given pump energy and crystal length,  we shall 
see  that the symmetric case $\eta=-1$ is much more efficient than the asymmetric case $|\eta| \ll 1$:  this is quite natural because in the first case both photons propagate close to the pump, while in the latter case one photonic wavepacket rapidly separates from it, so that the effective interaction length
is strongly reduced. \\
The probability of generating at least a  photon pair  can be easily calculated from  the state \eqref{psi} as
$
 \frac{\cal N}{1+{\cal N}} \approx {\cal N} $ for  ${\cal N} \ll 1$, 
where $ {\cal N} $ is the mean number of photon pairs per pulse. Within the linear approximation, making e.g. use of Eq. \eqref{psitemp}  we obtain 
\begin{align} 
{\cal N} &= \int d t_s \int d t_i \left |\phi(t_s,t_i )\right|^2  =g^2   \frac{\sqrt{\pi}\taup} {2 \Dtau}  
\end{align}
Notice that this formula is valid for any geometry and any value of $\taup$ (of course, provided that $\Dtau \ne 0$). However, the number of down-converted  pairs may be  quite different in different regimes. \\
In the  long-pump regime, where the state is highly entangled, the number of down-converted photons is approximately 
 \beq
{\cal N} \simeq \frac{g^2}{2}  \cappa \gg g^2, 
\eeq
 where we used  the asymptotic expression of  the Gaussian Schmidt number in Eq.\eqref{KGauss_long}].Roughly speaking, the number of photon pairs is  the number of photons per mode that one would have in the strictly CW pump regime, multiplied by the number of entangled modes $\cappa$. 
\\
Let us then consider the separable regimes.   In the symmetric case  $\eta =-1 $, 
\beq
{\cal  N}
 =\frac{ g^2 }{4}  \frac{\sqrt{\pi}  \taup} {|\taui |} \approx  \frac{ g^2 }{4}    \sqrt{ 2\pi \gamgauss} 
\eeq
where in the last expression  we inserted the Gaussian value of the pump duration \eqref{taupmin} 
$\taup^{min} 
=\sqrt{2\gamgauss|\taus\tauipm|}$ at which the best separability is achieved.  Considering that $  \sqrt{ 2\pi \gamgauss} \approx 1$, then the mean  number of down-converted photons is close to $g^2/4$.\\
In the asymmetric case  $\eta \ll 1$ (counterpropagating or co-propagating case with asymmetric group-velocity matching), instead 
\beq
{ \cal N}
= \frac{g^2 }{2}  \frac{\sqrt{\pi} \taup }{(1- \eta)|\tauipm| }  \simeq   \frac{g^2 }{2}   \frac{\sqrt{\pi}   \taup} { |\tauipm| }  \ll   
\frac{g^2 }{2}
\eeq
because separability is achieved for $\taup \ll |\tauipm| $. 
\footnote{If we use the Gaussian result for $\taup^{min} $, we also can write $   N \approx    \frac{  g^2}{2}      \sqrt{  |\eta| }     $ . }
Notice that in this latter case the efficiency is strongly reduced, because of the lack of superposition between the pump and the idler wave. We remind that this configuration is characterized by the fact that the idler wavepacket rapidly separates from the short pump pulse 
$\taup \ll |\tauipm|$. Since $g \propto l_c $ it effectively works as if the crystal length were reduced to a shorter length 
\beq
l_c \longrightarrow  l_c \sqrt{\frac{ \taup} { |\tauipm| }} .
\eeq
At this point one is naturally lead to wonder  whether filtering the spectral or temporal modes would not lead to the same or better efficiency.  If a  long  enough pump pulse is used , the number of photon pairs goes as $ \simeq \cappa \frac{g^2}{2} $.  Let us then suppose that it is possible to make a  "clever" filtering, which  projects the state onto a single Schmidt mode: even assuming that all the Schmidt modes have the same weight,  a fraction $1/\cappa$ of the total number of modes would then be  transmitted, leading to ${\cal N} = \frac{g^2}{2} $, which is anyway better than what could obtained with  asymmmetric group-velocity matching.  However, practical considerations, which are outside the scope of this theoretical work, may then lead to conclude that is anyway better to directly generate a factorable state

\section{Summary and conclusions}
In this work we compared different phase-matching configurations which are known to generate
 pure heralded single photons from spontaneous parametric down-conversion, namely the counter-propagating configuration, and the co-propagating configuration with  asymmetric and symmetric group velocity  matching.  
We provided a systematic 
analysis of the conditions under which uncorrelated  twin photons can be generated, which basically hold for any PDC configuration, provided that the bandwidth in play are not too broad.  In our description all the properties of the source are condensed in  two time scales $\taus, \tauipm$ characteristic of the relative propagation of the three waves inside the medium. \\
On the one side,  we performed a sort of standard analysis in the spectral domain, where we derived a simple formula for  the  Schmidt number of entanglement, that permits to evaluate the degree of purity of the heralded photons via a single parameter. 
 On the other side,  the less standard analysis of   the correlation of twin photons in the temporal domain clarifies  in a more physical sense the role of the two characteristic time scales, and permits to understand the mechanism under which the temporal correlation between twin photons can be eliminated. In particular it shows that one way of eliminating this temporal correlation relies on creating a  strong asymmetry between the velocities of the two down-converted waves relative to the pump wave, so that one photon  propagates locked under  the pump while the other,  for a  pump pulse  sufficiently localized inside the crystal, separates from it.   In this conditions,   the timing provided by the pump pulse may be more precise than that offered by detecting any of the twin photons, 
and  the arrival times of the two photons  appear uncorrelated. 
Such an asymmetry  in the relative propagation velocities is naturally present in the counter-propagating geometry, because of the natural separation of the GVM and the GVS time  scales, but requires particular tuning conditions in the co-propagating geometry. 
Because of this unique feature, counter-propagating twin photons in a pure state can in principle be heralded at any frequency 
by choosing the required poling period. Moreover, the twin photons
are naturally narrow-band, especially the one propagating opposite to the pump direction, and separability
is achieved for a broad range of pump pulse durations. 

An other way of eliminating the correlation, which  can be implemented only for co-propagating photons,   requires that the twin photons propagate symmetrically delayed with respect to the pump pulse.
Also in this case our analysis shows that in proper conditions 
the timing information which can be gained from a localized pump can be  better than from any of the twins, but the required pump duration is ultrashort, and the generated twins are broadband. 
Although the purity of heralded  photons generated in this way appears somehow lower, symmetric group velocity matching has the advantage of  a higher efficiency of pair production, because both twin photons propagate close to the pump. Conversely, the asymmetric group-velocity matching and the counterpropagating geometry are characterized by a low efficiency, because purity requires that one photonic wavepacket separates from the pump pulse, so that  the effective interaction length is reduced. 
\par 
In conclusion we provided a systematic study of the generation of  pure heralded single-photons through spontaneous PDC, which may turn useful for optimizing the existing configurations in view of different applications. For example when narrowband single-photons are required,  the counter-propagating configuration is the natural choice.  In addition,  once the technical challenges for the fabrication of crystals with sub-micrometer poling periods are overcome, this strategy offers more flexibility since it  can be virtually implemented for any wavelength, and generates pure single-photons for a broad range of pump durations. 
In contrast, twin photons emitted in the common co-propagating geometry are naturally broadband 
and can be generated in a separable state only for very short pulses, under particular tuning conditions.  Thus this is the good choice when broadband photons and or high repetition rates are required. Finally, if efficiency is the main issue, the symmetric group velocity matching should represent  the best choice. \\
We also hope that our general analysis, by providing a deeper and more intuitive understanding of the mechanism through which  the temporal entanglement of twin photons con be eliminated, may stimulate  new strategies of heralded photon generation.

\appendix
\section{Specific configurations} 
\label{sec:examples}
According to the results presented Sec. \ref{sec:temporal} and \ref{sec:Schmidt}, we considered  three  configurations  suitable for generating  pure heralded photons, and a specific example for each configurations : 

{\bf (i) Counter-propagating geometry ($|\eta|<<1$)}\\
The peculiarity of the counter-propagating geometry is  that the condition $|\eta|\ll 1$ is naturally fulfilled, so that any  phase-matching configuration  has the potentiality to generate pure heralded photons.  As a specific example, we considered 
a $10$mm long periodically poled crystal  of Potassium Titanyl Phosphate (PPKTP) in a type $0$ (e-ee) phase-matching configuration: the poling period is $\Lambda=800$nm, 
$\lambda_p=814.5$nm, $\lambda_s=1145$nm, $\lambda_i=2932.4$nm, $\eta=\taus/\taui=0.01$. 
Apart from the length of the crystal,
the parameters are those of  the experiment reported in \cite{canalias2007}), and are not particularly optimized for separability. Notice that for the same crystal,  the condition $\eta=0$ can be also realized, and would lead to a higher degree of purity, as discussed in Ref.\cite{gatti2015}.
\par
For the co-propagating geometry we considered  two examples of asymmetric $\eta=0$ and symmetric group velocity matching ($\eta=-1$), 
taken from the literature \cite{grice2001, uren2006, mosley2008, mosley2008b},  in particular the example (II) is that of the experiment 
by Mosley et al. \cite{mosley2008b}.
\par
{\bf (ii) Co-propagating geometry,  asymmetric group velocity matching  $\eta=0$}\\
We  considered a  $10$mm
   Potassium Dihydrogen Phosphate (KDP) crystal cut for type II collinear phase-matching (e-oe) at degeneracy.
When pumped at $415$nm with a tuning angle $\theta_p=68.7^\circ$ with the crystal axis, the KDP crystal
has the peculiarity of displaying
a vanishing GVM between pump and the signal field (i.e. $\taus=0$, $\eta=0$) and is therefore well
suited for the generation a separable two-photon state provided that 
$\taup\ll\tauib=0.72$ps \cite{grice2001,mosley2008}. 
\begin{figure*} 
\centering
\includegraphics[scale=0.55]{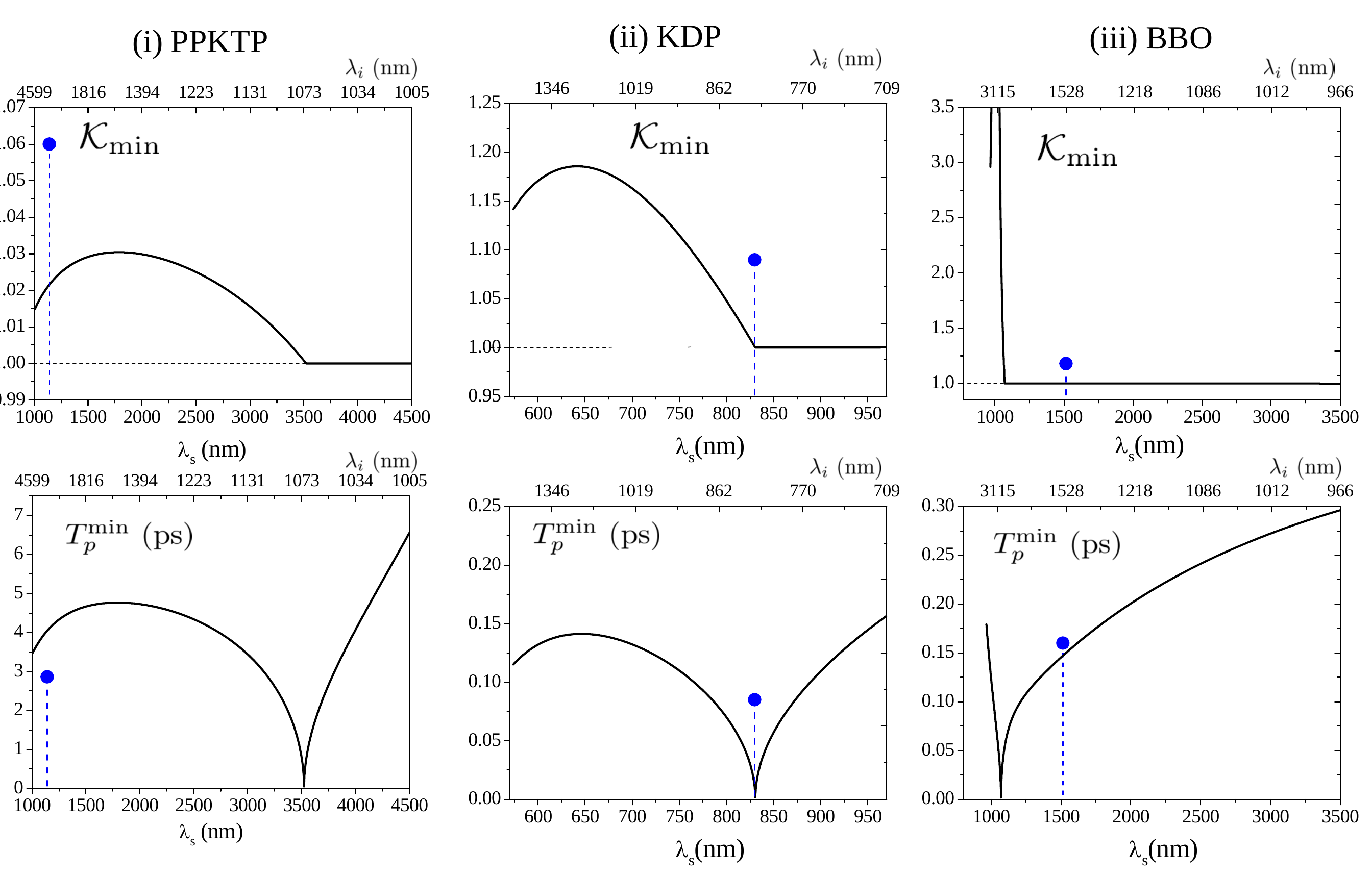}
\caption{Minimum of  the Schmidt number $K_{min}$ (top panels) and relative pump duration $\taup^{min}$ (bottom panels) as a function of the signal
wavelength,  evaluated with the Gaussian approximation in  Eqs.(\ref{Kmin}) and (\ref{taupmin}), for the three crystals in Table\ref{table1}. 
The  blue dots  are the value numerically calculated at the minima of $\cappa$ in Fig.\ref{fig3}, and correspond to the parameters  of Fig. \ref{fig4_temp}(b) (f) and (h). The top
horizontal scale shows the conjugate idler wavelength $\lambda_i$. }
\label{figKlambda}
\end{figure*}
\par
{\bf (iii) Co-propagating geometry, symmetric group velocity matching   $\eta=-1$}\\
The symmetric condition $\eta=-1$  can be fulfilled only in the co-propagating configuration,  and is rather difficult to meet because
it requires that the pump  inverse group velocity falls exactly midway between the signal and the idler inverse 
group velocities 
\beq
\tauib=-\taus\longleftrightarrow
\frac{1}{2}\left(\frac{1}{v_{gs}}+\frac{1}{v_{gi}}\right)=\frac{1}{v_{gp}}
\label{sym}
\eeq
Provided this relation is satisfied, the two-photon correlation $\psi(\Omega_s,\Omega_i)$  in the Gaussian approximation displays a circular shape 
for $\taup=\taup^{min}$, since $c_{12}=0$ and $c_{11}=c_{22}=2\gamma\taus^2$.
For the optimized pump pulse duration, the generated twin photons are thus not only uncorrelated but also indistinguishable.\\
As a specific example we considered a $10$mm  Beta-Barium Triborate (BBO) crystal
both cut for type II collinear phase-matching (e-oe) at degeneracy.
When pumped at $757$nm with a pump tuning angle $\theta_p=28.8^\circ$ 
the condition $\taus=-\taui=0.237ps$, $\eta=-1$ is realized. 
\par
Table \ref{table1} summarizes the parameters for three  examples chosen as 
representative of  the configurations {\bf (i)}, {\bf (ii)} and {{\bf (iii)}. 

Fig. \ref{figKlambda} plots  the results  of the Gaussian approximation for $K_{min}$ and $\taup^{min}$ [Eqs.  (\ref{Kmin}) and (\ref{taupmin})], as 
a function of the signal central wavelengths $\lambda_s$, for these three examples. The phase-matched wavelengths, $\lambda_s$ and  $\lambda_i$,
and the corresponding characteristic times $\taus$ and $\tauipm$ are evaluated using the Sellmeier dispersion formula
reported in \cite{niko91,kato2002,zernike64}.
For the PPKTP crystal, 
different wavelengths corresponds to different poling periods $\Lambda$, not reported in the figure.
For the bulk KDP and BBO crystals the signal and idler central wavelengths are varied by changing the tuning angle 
$\theta_p$ between the pump direction and the crystal axis (not reported in the figure). Notice that in
the BBO case $\eta$ is always negative for $\lambda_s>1070\,$nm, so that the generated two-photon state is separable 
when the crystal is tuned on those wavelengths according to approximation (\ref{Kmin}). 
Notice also that
for $\lambda_s=1010\,$nm the group velocities of the signal and idler fields becomes equal ($\eta=1$)
and the Schmidt number predicted by Eq.(\ref{Kgauss}) goes to infinity. 
Under these conditions the SPDC bandwidths and the number entangled modes are in fact very large, through not infinite,  as they are only limited by group-velocity dispersion, a feature not taken into account in the simplified model based 
the linearized propagation (\ref{Dlin}) (\ref{betalin}). 

---------------------------------\\

\bibliography{biblio_schmidt2017}
\bibliographystyle{apsrev4-1}
\end{document}